\documentclass[aps,preprint,showpacs,amssymb,eqsecnum,floatfix]{revtex4}
\usepackage{amsmath}
\usepackage{amsfonts}
\usepackage{amssymb}
\usepackage{graphicx}
\usepackage{bm}
\begin{document}
\title
{Minimum action method for the Kardar-Parisi-Zhang equation}
\author{Hans C. Fogedby}
\email{fogedby@phys.au.dk} \affiliation{Department of Physics and
Astronomy, University of Aarhus\\Ny Munkegade, 8000, Aarhus C, Denmark \\
and \\
Niels Bohr Institute\\
Blegdamsvej 17, 2100, Copenhagen {\O}, Denmark}
\author{Weiqing Ren}
\email{weiqing@cims.nyu.edu}
\affiliation{Courant Institute of
Mathematical Sciences, New York University\\251 Mercer Street, New
York, NY 10012, US}
\begin{abstract}
We apply a numerical minimum action method derived from the
Wentzell-Freidlin theory of large deviations to the
Kardar-Parisi-Zhang equation for a growing interface. In one
dimension we find that the switching scenario is determined by the
nucleation and subsequent propagation of facets or steps,
corresponding to moving domain walls or growth modes in the
underlying noise driven Burgers equation. The transition scenario
is in accordance with recent analytical studies of the one
dimensional Kardar-Parisi-Zhang equation in the asymptotic weak
noise limit. We also briefly discuss transitions in two
dimensions.
\end{abstract}

\maketitle
\section{\label{intro}Introduction}
The large majority of natural phenomena are characterized by being
out of equilibrium. This class includes turbulence in fluids,
interface and growth problems, chemical reactions, processes in
glasses and amorphous systems, biological processes, and even
aspects of economical and sociological structures
\cite{Nelson03,Chaikin95}. In this context there is a continuing
interest in the strong coupling aspects of stochastically driven
non equilibrium model systems \cite{Cross94,Barabasi95}. Here the
dynamics of complex systems driven by weak noise, corresponding to
rare events, is of particular interest. The issue of different
time scales characterizes many interesting processes in nature.
For instance, in the case of chemical reactions the reaction time
is often orders of magnitude larger than the molecular vibration
periods \cite{Geissler01}. The time scale separation problem is
also encountered in the case of conformational changes of
biomolecules \cite{Olender96}, nucleation events during phase
transitions, switching of the magnetization in magnetic materials
\cite{Berkov98,Koch00}, and even in the case of comets exhibiting
rapid transitions between heliocentric orbits around Jupiter
\cite{Jaffe02}.

In the weak noise limit the standard Monte Carlo method or direct
simulation of the Langevin equation becomes impractical owing to the
large separation of time scales and alternative methods have been
developed. The most notable analytical approach is the formulation
due to Freidlin and Wentzel which yields the transition
probabilities in terms of an action functional \cite{Freidlin98}.
This approach is the analogue of the variational principle proposed
by Machlup and Onsager \cite{Machlup53,Onsager53}, see also work by
Graham et al. \cite{Graham84,Graham90} and Dykman \cite{Dykman90}.
The Freidlin-Wentzel approach is also equivalent to the
Martin-Siggia-Rose method \cite{Martin73} in the weak noise limit of
the path integral formulation \cite{deDominicis75,deDominicis76,
Baussch76,Janssen76,deDominicis78}. In order to overcome the time
scale gap various numerical methods have also been proposed. We
mention here the transition path sampling method \cite{Bolhuis02}
and the string method \cite{Ren02,E02a,E05,Ren05}.

A particularly interesting non equilibrium problem of relevance in
the nanophysics of magnetic switches is the influence of thermal
noise on two-level systems with spatial degrees of freedom
\cite{Berkov98,Koch00,Garcia99}. In a recent paper by E, Ren, and
Vanden-Eijnden \cite{E04}, see also Ref. \cite{E03}, this problem
has been addressed using the Ginzburg-Landau (GL) equation driven
by thermal noise. Applying the field theoretic version of the
Onsager-Machlup functional \cite{Onsager53,Machlup53} in the
Freidlin-Wentzell formulation \cite{Freidlin98},
these authors developed the so called minimum action method
in which they implemented a powerful
numerical optimization techniques for the determination of the
space-time configuration which minimizes the Freidlin-Wentzell action.
The minimizers correspond to the kinetic pathways and the
associated action yields the switching probabilities in the long
time-low temperature limit. In the picture emerging from the
numerical study the switching between metastable states
is due to noise induced nucleation and subsequent propagation of
domain walls across the sample. Subsequently, we supplemented the
work by E et al. and presented a dynamical description and
analysis of the non equilibrium transitions in the noisy 1D GL
equation for an extensive system based on a weak noise canonical
phase space formulation of the Freidlin-Wentzel or
Martin-Siggia-Rose methods \cite{Fogedby03a,Fogedby04b}. We
derived propagating nonlinear domain wall or soliton solutions of
the resulting canonical field equations with superimposed
diffusive modes. The transition pathways are characterized by the
nucleations and subsequent propagation of domain walls. We
discussed the general switching scenario in terms of a dilute gas
of propagating domain walls and evaluated the Arrhenius factor in
terms of the associated action. In conclusion we found excellent
agreement with the numerical studies by E et al.
\cite{E03,E04}.

The noise-driven GL equation belongs to the class of so-called
gradient systems where the drift term in the Langevin equation can
be derived from a free energy functional. Regarding the kinetic
transitions this property implies the existence of an underlying
free energy landscape in which the optimal pathway proceeds via
saddle points, yielding the corresponding Arrhenius factor. We
note that this interpretation implies a fluctuation-dissipation
theorem relating the strength of the noise to the kinetic
transport coefficient. There is, however, another interesting
class of stochastic model systems characterized by Langevin
equations, where the drift term cannot be associated with a free
energy functional. Those are the so called non-gradient systems
for which the interpretation of pathways in a free energy
landscape fails and has to be replaced by pathways in an ``action
landscape''. In recent work, see e.g. Refs.
\cite{Fogedby05a,Fogedby06a}, where earlier references can be
found, we have addressed a non equilibrium model falling in the
class of non-gradient systems, namely the Kardar-Parisi-Zhang
(KPZ) equation or, equivalently, noisy Burgers equation
describing, for example, a growing interface. Using the weak noise
canonical phase space method alluded to above, we find that the
kinetic pathways correspond to nucleation and propagation of
localized growth modes with superimposed diffusive modes. The
growth modes together with the diffusive modes carry an action,
yielding the transition probabilities.

The purpose of the present paper is to attempt to substantiate the
weak noise growth mode approach to the KPZ equation by a direct
numerical optimization employing the minimum action method developed
by E, Ren, and Vanden-Eijnden \cite{E04}, see also Refs.
\cite{Zhou08,Heymann08}. Similar to the GL case we find in 1D that
the switching scenario is determined by the nucleation and
propagation of growth modes. We are also able to account numerically
for the associated transition probabilities. The paper is organized
in the following manner. In Sec. \ref{kpze} we briefly review the
KPZ equation and the analytical results obtained by the weak noise
canonical phase space approach. In Sec. \ref{mam} we introduce the
minimum action method and establish the connection with the phase
space method and path integral formulations. In Sec. \ref{num} we
discuss the numerical implementation of the Freidlin-Wentzel scheme.
In Sec. \ref{numres} we present the numerical results for various
switching scenarios. In Sec. \ref{disc} we offer a heuristic
discussion of the numerical data based on the analytical phase space
method. In Sec. \ref{2d} we briefly discuss kinetic pathways in 2D
and present some switching scenarios. Sec. \ref{sum} is devoted to a
summary and a conclusion.
\section{\label{kpze}The KPZ equation}
In this section we review the KPZ equation and apply the weak
noise method. The KPZ equation describes an intrinsic non
equilibrium problem and plays in some sense the same role as the
Ginzburg-Landau functional in equilibrium physics
\cite{Chaikin95,Binney92}. The KPZ equation was introduced in 1986
in a seminal paper by Kardar, Parisi, and Zhang
\cite{Kardar86}, see also Ref. \cite{Medina89,Halpin95}, and
purports to describe non equilibrium aspects of a growing
interface \cite{Krug97,Barabasi95}. In a Monge representation the
KPZ equation for the stochastic time evolution of the height field
$h({\bf r},t)$ has the form
\begin{eqnarray}
&&\frac{\partial h}{\partial t}=\nu{\bm\nabla}^2h
+\frac{\lambda}{2}({\bm\nabla} h)^2-F+\eta, \label{kpz}
\\
&&\langle\eta\eta\rangle({\bf r},t)=\Delta\delta^d({\bf
r})\delta(t). \label{kpznoise}
\end{eqnarray}
Here the damping coefficient or viscosity $\nu$ characterizes the
linear diffusion term $\nu\nabla^2h$, the growth parameter
$\lambda$ controls the strength of the nonlinear growth term
$(\lambda/2)({\bm\nabla} h)^2$, $F$ is the imposed drift
term and is a constant here, and $\eta$ a locally correlated
white Gaussian noise,
modelling the stochastic nature of the drive or environment; the
noise correlations are characterized by the noise strength
$\Delta$.
\subsection{Burgers and  Cole-Hopf equations}
In the growth mode analysis of the KPZ equation the local slope of
the growing interface given by the vector field,
\begin{eqnarray}
{\bf u}={\bm\nabla}h, \label{slope}
\end{eqnarray}
is of importance. In terms of ${\bf u}$ the KPZ equation then maps
onto the Burgers equation driven by conserved noise
\cite{Forster76,Forster77,E00, Woyczynski98}
\begin{eqnarray}
\frac{\partial{\bf u}}{\partial t} =\nu{\bm\nabla}^2{\bf u}+
\lambda({\bf u}\cdot{\bm\nabla}){\bf u}+{\bm\nabla}\eta.
\label{bur}
\end{eqnarray}
In the deterministic case for $\eta=0$ the Burgers equation has
been used to study irrotational fluid motion and turbulence
\cite{Burgers29,Burgers74,Saffman68,Jackson90,Whitham74}; it has
also played a role in astrophysics in the context of large scale
structures in the universe \cite{Zeldovitch72,Shandarin89}.

Another quantity of importance in our analysis of the KPZ equation
is the diffusive field $w$ defined by the non-linear Cole-Hopf
transformation \cite{Cole51,Hopf50,Medina89}
\begin{eqnarray}
w=\exp\left[\frac{\lambda}{2\nu}h\right]. \label{ch}
\end{eqnarray}
In terms of $w$ the KPZ equation maps onto a linear diffusion
equation driven by a multiplicative noise, here denoted the
Cole-Hopf (CH) equation,
\begin{eqnarray}
\frac{\partial w}{\partial t} = \nu{\bm\nabla}^2w
-\frac{\lambda}{2\nu}w F+\frac{\lambda}{2\nu}w\eta. \label{che}
\end{eqnarray}
In the absence of noise for $\eta=0$ the CH equation reduces to
the linear diffusion equation and is readily analyzed permitting a
complete discussion of the KPZ and Burgers equations in the
deterministic case \cite{Medina89,Kardar86}. In the noisy case a
path integral representation maps the CH equation and consequently
the KPZ equation onto a model of a directed polymer in a quenched
random potential. The disordered directed polymer constitutes a
toy model within the spin glass literature and has been analyzed
by means of replica, Bethe ansatz, and functional renormalization
group techniques \cite{Kardar87a,Halpin95,LeDoussal05,Halpin98}.
\subsection{Scaling properties}
Most work on the KPZ equation has addressed the scaling issues.
For completeness we summarize the salient features here. The KPZ
equation lives at a critical point and conforms to the dynamical
scaling hypothesis \cite{Barabasi95,Family86,Family90,Family85}
which in terms of the height correlations assumes the form:
\begin{eqnarray}
\langle h h\rangle({\bf r},t) = r^{2\zeta}f(t/r^z). \label{scal}
\end{eqnarray}
Here $\zeta$ is the roughness exponent, $z$ the dynamical
exponent, and $f$ the associated scaling function. The exponent
$\zeta$ is a measure of the roughness of the interface, e.g., for
$\zeta=0$ the interface is flat, for $\zeta=1/2$ the interface
exhibits a random walk profile, $\langle hh\rangle({\bf r})\sim
r$. The exponent $z$ is a measure of the dynamical scaling, e.g.,
for diffusive behavior $z$ locks onto 2; for a 1D growing
interface $z=3/2$.

\begin{figure}[h!]
\includegraphics[width=1.0\hsize]{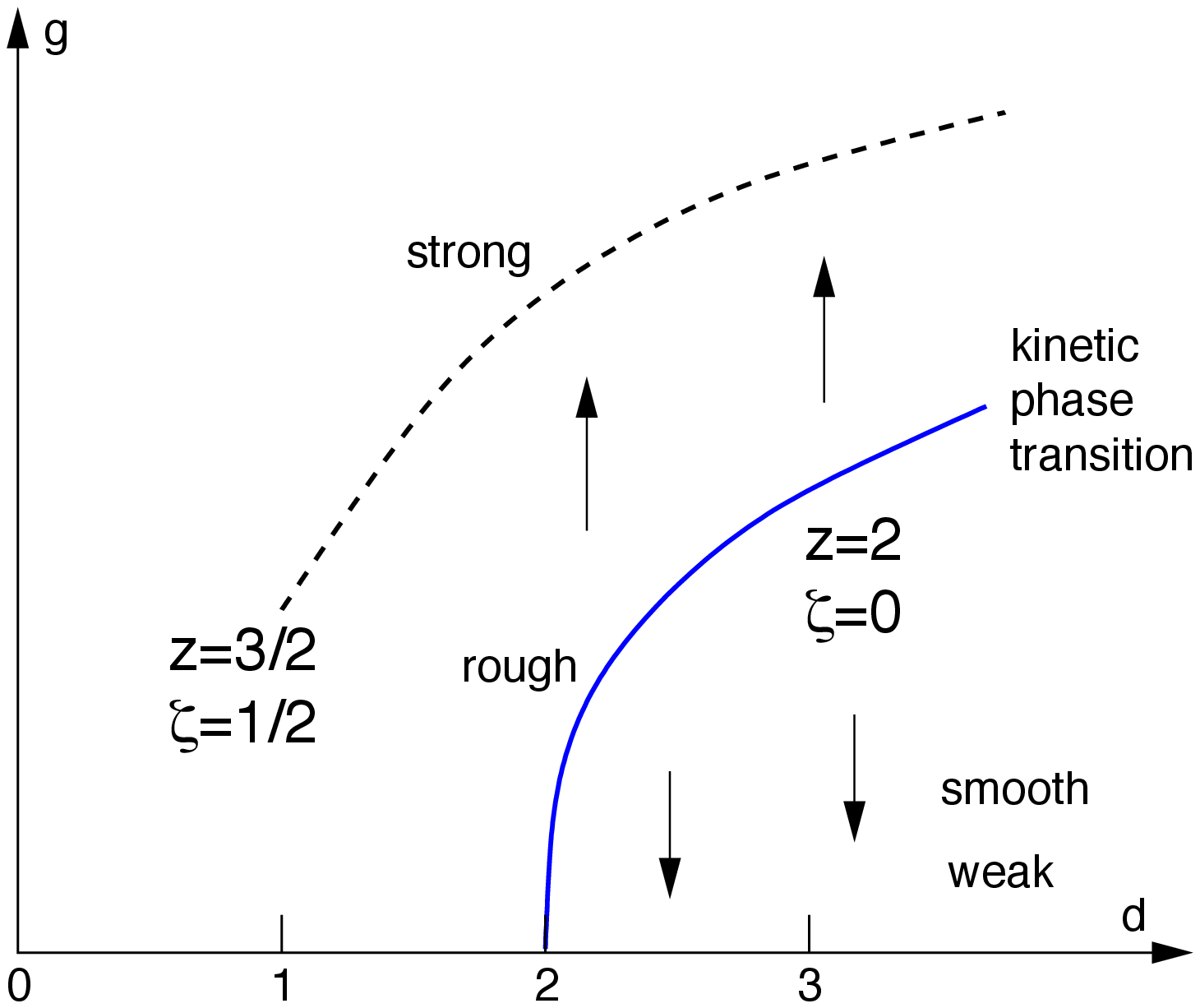}
\caption{DRG phase diagram for the KPZ equation to leading loop
order in $d-2$. We plot the effective coupling strength
$g=\lambda^2\Delta/\nu^3$ as a function of the dimension $d$. In
$d=1$ the DRG flow is towards a strong coupling fixed point with
scaling exponents $\zeta=1/2, z=3/2$. Above the lower critical
dimension $d=2$ there is a kinetic transition line, delimiting a
rough phase from a smooth phase. On the phase line $z=2$ and
$\zeta=0$. The weak coupling smooth phase is characterized by
$z=2$ and $\zeta=(2-d)/2$. Above $d=1$ the scaling exponents in
the strong coupling rough phase are unknown} \label{fig1}
\end{figure}

In order to extract scaling properties the initial analysis of the
KPZ equation was based on the dynamic renormalization group (DRG)
method, previously developed and applied to dynamical critical
phenomena and noise driven hydrodynamics
\cite{Forster76,Forster77,Hohenberg77}. An expansion in powers of
$\lambda$ in combination with a momentum shell integration yields
to leading order in $d-2$ the DRG equation $dg/dl=\beta(g)$, with
beta-function $\beta(g)=(2-d)g+\text{const.} g^2$
\cite{Wiese97,Wiese98}. Here $g=\lambda^2\Delta/\nu^3$ is the
effective coupling strength and $l$ the logarithmic scale
parameter. Below the lower critical dimension $d=2$ the DRG flow
is towards a strong coupling fixed point with scaling exponents
$\zeta=1/2$ and $z=3/2$ in $d=1$. Above $d=2$ a kinetic phase
transition line delimits a strong coupling regime from a weak
coupling regime. In the strong coupling regime the DRG flow is
towards a still poorly understood strong coupling fixed point with
unknown scaling exponents and scaling function. In the weak
coupling regime the DRG flow is towards a weak coupling fixed
point with scaling exponents $z=2$ and $\zeta = (2-d)/2$. On the
transition line $z=2$ and $\zeta=0$, see e.g. Ref.
\cite{Fogedby06a}. In Fig.~\ref{fig1} we have depicted the scaling
properties in a plot of the coupling strength $g$ versus the
dimension $d$.

We note two further properties of the KPZ equation which are also
relevant in a scaling context. Firstly, subject to a Galilean
transformation the equation is invariant provided we add a
constant slope to the height field $h$ and adjust the drift $F$
accordingly, i.e.,
\begin{eqnarray}
&&{\bf r}\rightarrow{\bf r}-\lambda{\bf u}^0t, \label{gal1}
\\
&&h\rightarrow h+{\bf u}^0\cdot{\bf r}, \label{gal2}
\\
&&F\rightarrow F+(\lambda/2){\bf u}^0\cdot{\bf u}^0; \label{gal3}
\end{eqnarray}
note that the slope field ${\bf u}$ and diffusive field $w$
transform according to ${\bf u}\rightarrow{\bf u}+{\bf u}^0$ and
$w\rightarrow w\exp[(\lambda/2\nu){\bf u}^0\cdot{\bf r}]$,
respectively. From a simple scaling argument and also following from
the DRG analysis the Galilean invariance implies the scaling law
\begin{eqnarray}
\zeta+z=2, \label{scallaw}
\end{eqnarray}
relating the roughness and dynamic scaling exponents. The Galilean
invariance is a fundamental dynamical symmetry specific to the KPZ
equation, delimiting the KPZ universality class. Secondly, a
fluctuation dissipation theorem is operational in 1D in the sense
that the stationary Fokker-Planck equation associated with the KPZ
equation admits the explicit solution \cite{Huse85,Halpin95}
\begin{eqnarray}
P_0(h)\propto\exp\left[-\frac{\nu}{\Delta}\int dx (\nabla
h)^2\right]. \label{stat}
\end{eqnarray}
The Gaussian form of the distribution shows that the slope
$u=\nabla h$ fluctuations are uncorrelated and that the height
field $h=\int^x u dx'$ performs a random walk in $x$. Note also
that the distribution is independent of the non-linear growth
parameter $\lambda$.
\subsection{Weak noise method}
Whereas the DRG approach, based on an asymptotic expansion about
the critical dimension $d=2$, deals with the long time - large
distance scaling properties of the KPZ equation, the asymptotic
weak noise approach addresses the stochastic growth morphology or
many body aspects. The weak noise or canonical phase space method
focuses on the noise strength $\Delta$ as the relevant parameter
in the problem. In the absence of noise for $\eta=0$ or $\Delta=0$
the morphology of the deterministic KPZ equation decays subject to
a transient pattern formation; in 1D corresponding to cusps
connected by parabolic segments \cite{Medina89}. In the presence
of even weak noise the KPZ equation is eventually driven into a
stationary stochastic state; the cross-over time diverging in the
limit of vanishing noise. In this sense the noise strength
$\Delta$ is a singular parameter and a weak noise approach
asymptotic in $\Delta$.

The weak noise canonical phase space approach is implemented by
applying an eikonal or WKB approximation to the Fokker-Planck
equation associated with the Langevin equation. Viewing the
Fokker-Planck equation as an imaginary time Sch\"odinger equation
the scheme is equivalent to the well-known WKB or semi-classical
approximation in quantum mechanics, where the wave function $\Psi$
is related to the classical action $S$ by
$\Psi\propto\exp[iS/\hbar]$, $\hbar$ being the Planck constant. In
quantum mechanics the quantum fluctuations characterized by
orbitals are then in the correspondence limit $\hbar\rightarrow 0$
replaced by orbits as solutions to the classical equations of
motion following from the action $S$.

In the weak noise approach the point of departure is a general
Langevin equation of the form
\begin{eqnarray}
&&\frac{dx}{dt}=-F(x)+\eta(t), \label{lan}
\\
&&\langle\eta\eta\rangle(t)=\Delta\delta(t); \label{noise}
\end{eqnarray}
for simplicity we consider a single random variable $x(t)$; for
the more general case see e.g. Refs.
\cite{Stratonovich63,Fogedby06a}. Here $F(x)$ is a general
non-linear drift, and $\eta$ an additive white noise correlated with
strength $\Delta$. In order to implement the weak noise
approximation we consider the equivalent Fokker-Planck equation
for the distribution $P(x,t)$:
\begin{eqnarray}
\Delta\frac{\partial P}{\partial t}=
\frac{1}{2}\Delta^2\frac{\partial^2P}{\partial
x^2}+\Delta\frac{\partial}{\partial x}(FP). \label{fp}
\end{eqnarray}
Interpreting $\Delta\partial/\partial x$ as a momentum operator,
$P$ as an effective wave function, and $\Delta$ as an effective
Planck constant, Eq. (\ref{fp}) has the form of an imaginary time
Schr\"odinger equation. Consequently, in the weak noise limit it
is suggestive to introduce the WKB or eikonal approximation
\cite{Landau59c}
\begin{eqnarray}
P(x,T)\propto\exp\left[-\frac{S(x,T)}{\Delta}\right]. \label{wkb}
\end{eqnarray}
To leading order in $\Delta$ the action $S$ then obeys a principle
of least action $\delta S=0$ as expressed by the Hamilton-Jacobi
equation $\partial S/\partial t+H(x,p)=0$ with associated
canonical momentum $p=\partial S/\partial x$
\cite{Goldstein80,Landau59b}. The Hamiltonian (energy) takes the
form
\begin{eqnarray}
H=\frac{1}{2} p^2-pF=\frac{1}{2}p(p-2F), \label{ham}
\end{eqnarray}
yielding the coupled Hamilton equations of motion
\begin{eqnarray}
&&\frac{dx}{dt}=-F+p, \label{eq1}
\\
&&\frac{dp}{dt}=p\frac{dF}{dx}.\label{eq2}
\end{eqnarray}
Finally, the action associated with an orbit from $x_i$ to $x$ in
the transition time $T$ is given by
\begin{eqnarray}
S(x,T)= \int_{x_i,0}^{x,T} dt\left[p\frac{dx}{dt} - H\right],
\label{act1}
\end{eqnarray}
or inserting the equations of motion for $x$
\begin{eqnarray}
S(x,T)= \frac{1}{2}\int_{x_i,0}^{x,T} dt~p^2. \label{act2}
\end{eqnarray}
The issue of solving the stochastic Langevin equation (\ref{lan})
or, equivalently, the deterministic Fokker-Planck equation
(\ref{fp}) in the weak noise limit $\Delta\rightarrow 0$ is then
replaced by, as  first step, solving the coupled equations of
motion (\ref{eq1}-\ref{eq2}) for an orbit from an initial
configuration $x_i$ at time $t=0$ to a final configuration $x$ at
time $t=T$. In the next step we evaluate the action $S$ associated
with the orbit and infer from the WKB ansatz (\ref{wkb}) the
transition probability for the specific transition. We note that
the noise in Eq. (\ref{lan}) has been replaced by the canonical
momentum $p$ and that $p$ is a dependent variable which has to be
chosen in accordance with the initial and final values of $x$ and
the imposed transition time $T$.

\begin{figure}
\includegraphics[width=1.0\hsize]{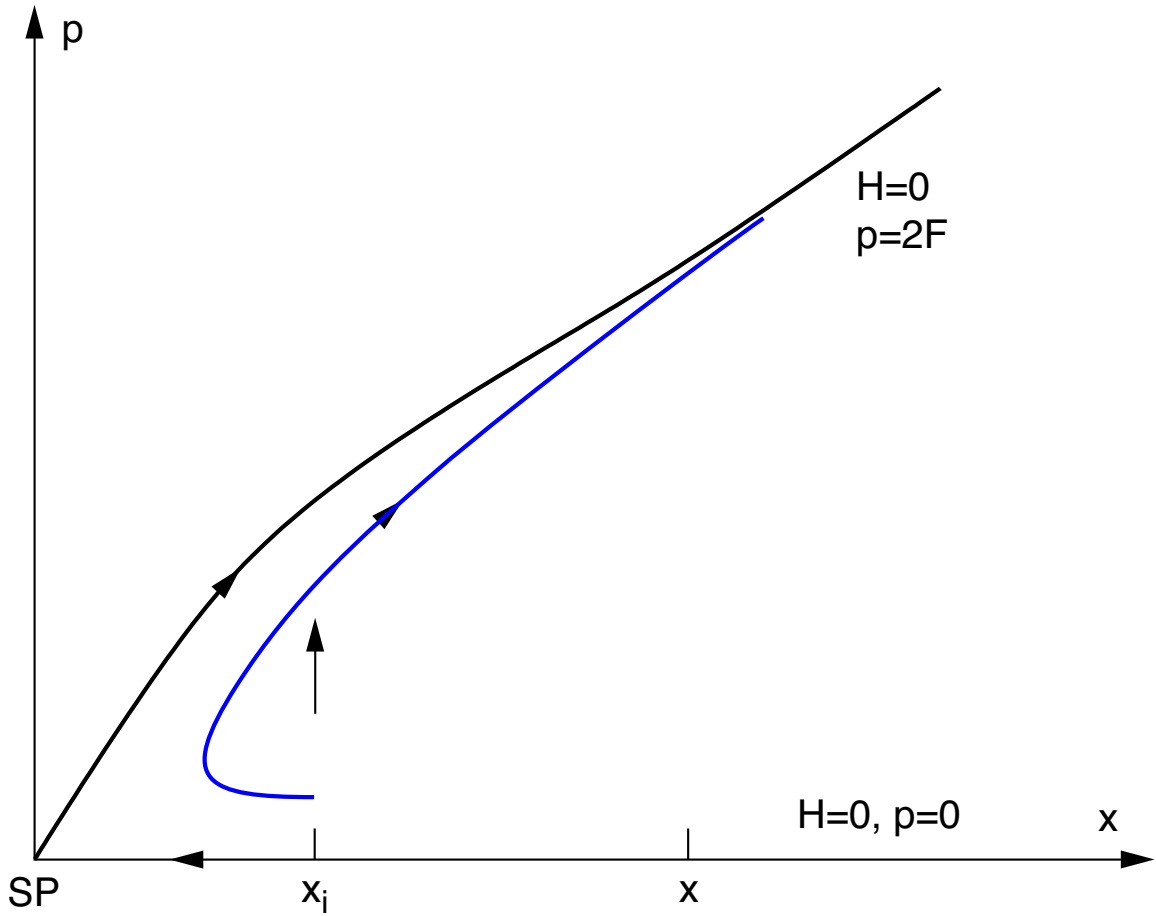}
\caption{Phase space representation of the weak noise method. We
show the zero energy submanifold $p=0$ corresponding to the
transient behavior and the submanifold $p=2F$ determining the
stationary distribution. The manifolds intersect in a hyperbolic
saddle point (SP). The infinite waiting time at SP corresponds to
the long time Markoff behavior. We depict a finite time orbit from
$x_i$ to $x$ in transition time $T$ and an infinite time orbit
passing through the saddle point.}\label{fig2}
\end{figure}

In a phase space representation the zero-energy manifolds $p=0$
and $p=2F$ intersecting at a hyperbolic saddle point play an
important role in determining the long time stationary
distribution $P_0(x)= \lim_{T\rightarrow\infty}P(x,T)$. Initially
an orbit from $x_i$ to $x$ moves along the the transient
zero-energy manifold $p=0$ towards the saddle point. This part of
the orbit represents the transient motion. As time progresses the
orbits bends away from the saddle point and is attracted to the
stationary submanifold $p=2F$. This part of the orbit corresponds
to the cross-over to a stationary random motion. In the limit of a
long transition time the orbit from $x_i$ to $x$ passes close to
the saddle point, at which the large waiting time ensures the Markoff
property. In Fig.~\ref{fig2} we have sketched the $\{x,p\}$ phase
space showing the zero-energy submanifolds, the saddle point and
an orbit from $x_i$ to $x$ in transition time $T$.
\subsection{Growth modes}
In the KPZ case the weak noise scheme is most easily implemented
for the CH equation (\ref{che}) driven by multiplicative noise.
This requires an extension of the weak noise method discussed in
Ref. \cite{Fogedby05a,Fogedby06a}. Introducing the wavenumber
parameters
\begin{eqnarray}
&&k= (\lambda F/2\nu^2)^{1/2}, \label{par1}
\\
&&k_0=\lambda/2\nu, \label{par2}
\end{eqnarray}
setting inverse length scales, we find the weak noise Hamiltonian
\begin{eqnarray}
H=\int d^dx~\left(p[\nu\nabla^2-\nu k^2]w+(1/2)k_0^2(wp)^2\right),
\label{hamch}
\end{eqnarray}
and associated field equations
\begin{eqnarray}
&&\frac{\partial w}{\partial t}=\nu[\nabla^2 w-k^2 w]+k_0^2w^2p,
\label{eq11}
\\
&&\frac{\partial p}{\partial t}=-\nu[\nabla^2 p-k^2 p]-k_0^2p^2w,
\label{eq22}
\end{eqnarray}
determining orbits in a $\{w,p\}$ phase space. Likewise, one
infers the action
\begin{eqnarray}
S(w,T)=\frac{1}{2}k_0^2\int^{w,T} d^dx dt (wp)^2, \label{actw}
\end{eqnarray}
yielding the transition probability to leading asymptotic order in
$\Delta$
\begin{eqnarray}
P(w,T)\propto\exp\left[-\frac{S(w,T)}{\Delta}\right]; \label{disw}
\end{eqnarray}
note that on the  $p=0$ manifold Eq. (\ref{eq11}) reduces to the
deterministic CH equation for $\eta=0$.

The equations of motion (\ref{eq11}-\ref{eq22}) serve two
purposes. On the one hand, a solution or orbit in phase space from
an initial configuration $w_i({\bf r})$ at time $t=0$ to a final
configuration $w({\bf r})$ at time $t=T$ with $p$ as an adjusted
noise field yields an action $S$ and thus a contribution to the
transition probability $P(w,T)$. Secondly, the solution $w({\bf
r},t)$ interpreted as a classical orbit also provides a growth
morphology for the CH equation. The deterministic growth or
evolution of the diffusive field $w$ then corresponds to a growth
morphology for the KPZ equation by means of the inverse Cole-Hopf
transformation
\begin{eqnarray}
h=(1/k_0)\ln w. \label{ch2}
\end{eqnarray}
Likewise, the transition probability $P(h,T)$ is given by
\begin{eqnarray}
P(h,T)=\int\prod_{\bf r}dw\delta(h-(1/k_0)\log w)P(w,T).
\label{dish}
\end{eqnarray}
The growth morphology follows from the coupled nonlinear field
equations (\ref{eq11}-\ref{eq22}). Owing to the negative diffusion
coefficient the equations are numerically unstable, see Ref.
\cite{Fogedby02a}, however, searching for localized instanton or
soliton type solutions we note that on the $p=0$ and $p=\nu w$
submanifolds the static equations reduce to the static diffusion
equation and the static nonlinear Schr\"odinger equation, well
known in the context of dark solitons in Bose condensed atomic
gasses \cite{Fogedby02a}
\begin{eqnarray}
&&\nabla^2 w=k^2 w, \label{diff}
\\
&&\nabla^2 w=k^2 w-k_0^2w^3. \label{nls}
\end{eqnarray}
%
\subsection{Domain walls in 1D}
\begin{figure}
\includegraphics[width=1.0\hsize]{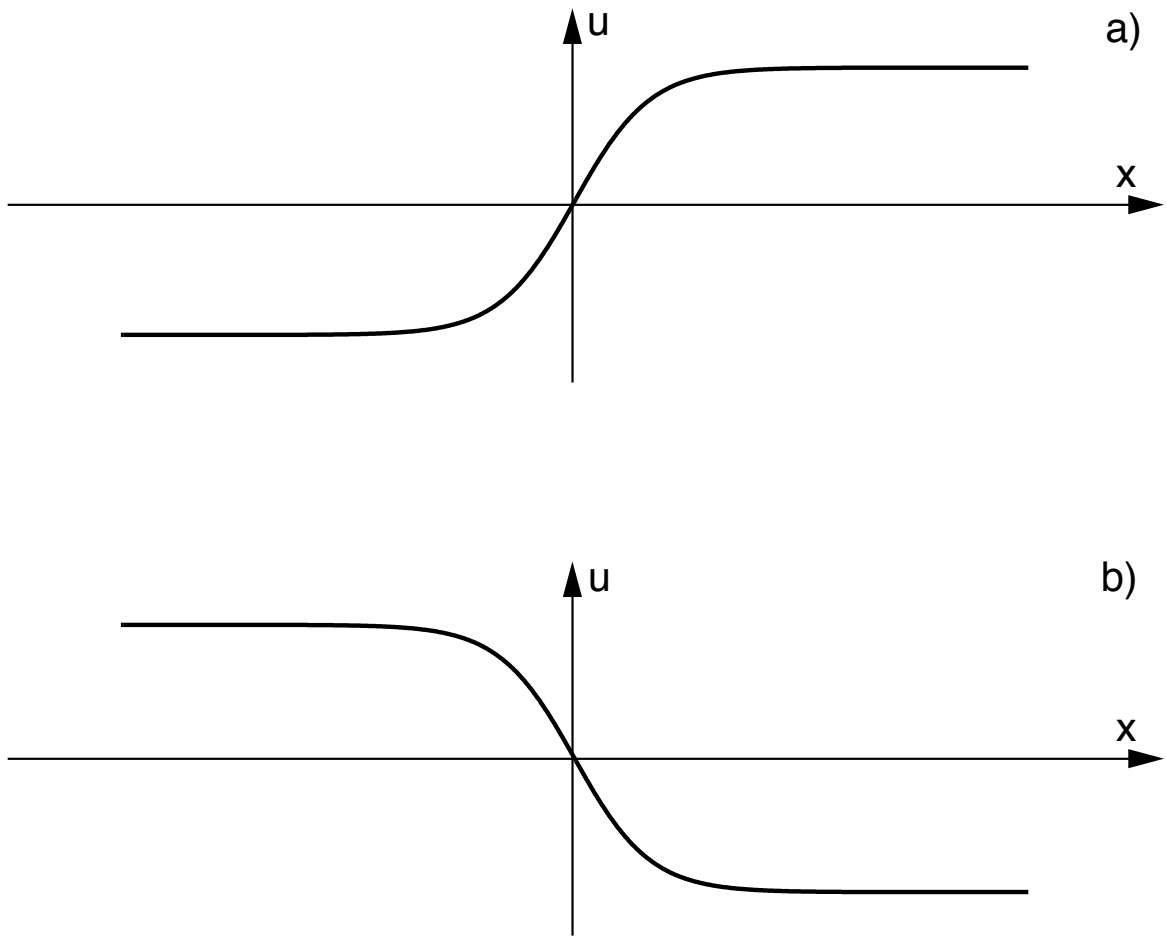}
\caption{We depict the static domain walls in the slope field
corresponding to the solutions of the diffusion and non linear
Schr\"odinger equations for the diffusive field . In a) we show
the right hand domain wall. This domain wall carries vanishing
action and is identical to the viscosity smoothed shock waves of
the noiseless Burgers equation. In b) we show the noise induced
left hand domain wall carrying a finite action.} \label{fig3}
\end{figure}

In 1D Eqs. (\ref{diff}-\ref{nls}) admit the static solutions
$w_\pm\propto\cosh^{\pm 1} kx$ for the diffusive field $w$. These
modes correspond to cusps in the height field,
$h_\pm=\pm(1/k_0)\ln (\cosh kx)$, and to static domain walls or
solitons in the local slope field:
\begin{eqnarray}
u_\pm(x)=\pm \frac{k}{k_0}\tanh kx. \label{dom}
\end{eqnarray}
The right hand domain wall, $u_+(x)=(k/k_0)\tanh kx$, is
associated with the $p=0$ manifold and carries zero energy and
zero action. This mode is the well-known viscosity-smoothed shock
wave solution of the static noiseless Burgers equation
$\nu\nabla^2 u+\lambda u\nabla u=0$, as easily seen by inspection
\cite{Woyczynski98}. The left hand domain wall, $u_-(x)=
-(k/k_0)\tanh kx$, lives on the $p=\nu w$ manifold and carries a
finite action
\begin{eqnarray}
S=\frac{8\nu^2k^3}{3k_0^2}T; \label{actdw}
\end{eqnarray}
the static domain walls are depicted in Fig.~\ref{fig3}. By
applying the Galilean transformation (\ref{gal1}-\ref{gal3}) the
static domain walls can be boosted to a finite propagation
velocity and we obtain the moving domain walls or growth modes
\begin{eqnarray}
u_\pm(x,t)=\pm \frac{k}{k_0}\tanh k(x-\lambda u^0t)-u_0.
\label{movdom}
\end{eqnarray}
The propagating domain walls form the basic building blocks in the
construction of a growth morphology. Considering a dilute gas of
non overlapping growth modes of different amplitudes or ``charges''
$k_i$, where a positive charge corresponds to a right hand domain
wall and a negative charge to a left hand domain wall, we obtain
the global solution \cite{Fogedby06a}
\begin{eqnarray}
&&u(x,t)= \frac{1}{k_0}\sum_i k_i\tanh
|k_i|(x-x_i(t)),~~~~~~~~~~~~ \label{sol1}
\\
&&h(x,t)= \frac{1}{k_0}\sum_i\frac{k_i}{|k_i|}
\ln(\cosh|k_i|(x-x_i(t))),  \label{sol2}
\\
&&x_i(t)=\int_0^tv_i(t')dt'+x_i^0, \label{sol3}
\\
&&v_i(t)=-\frac{\lambda}{k_0}\sum_{l\neq i}k_l\tanh
|k_i|(x_i(t)-x_l(t));\label{sol4}
\end{eqnarray}
note that the neutrality condition $\sum_ik_i=0$ ensures that the
interface is flat at infinity. This condition, however, allows for
an offset in $h$ corresponding to propagating facets.

\begin{figure}
\includegraphics[width=1.0\hsize]{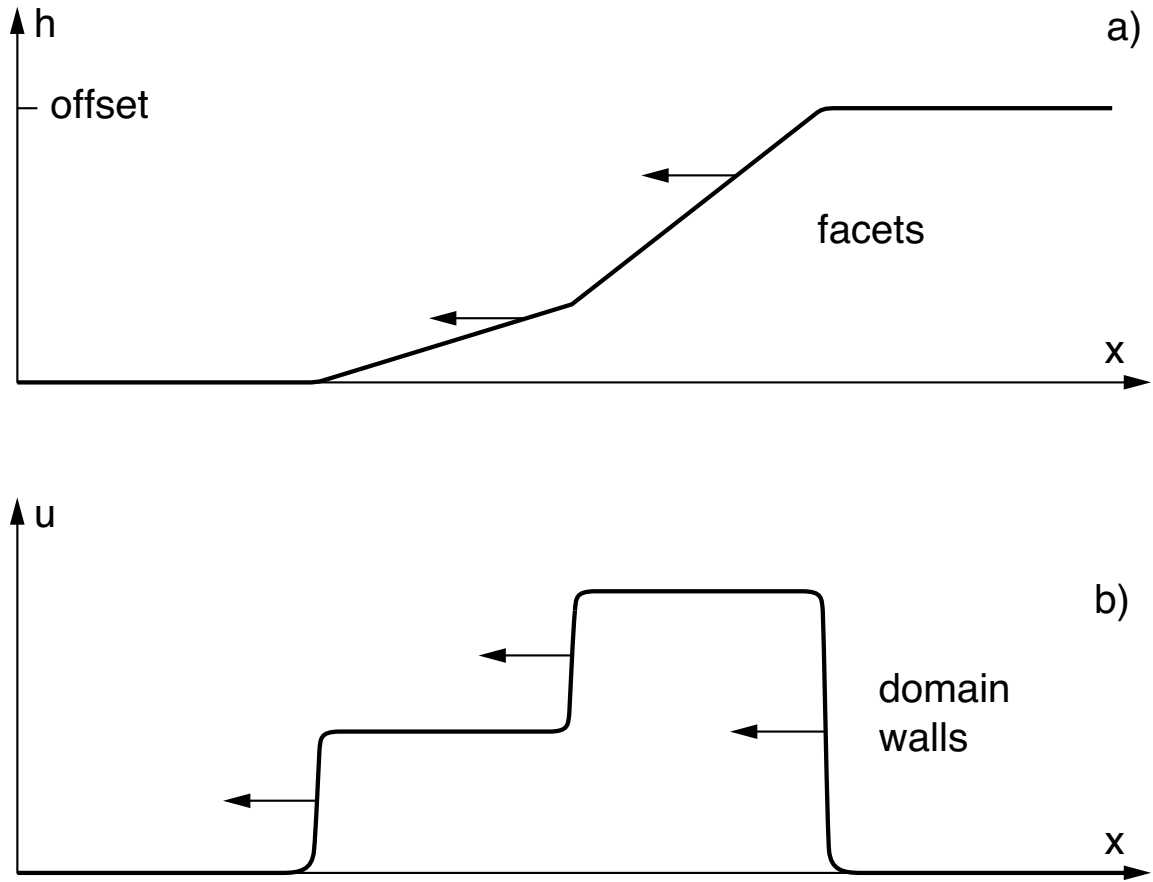}
\caption{We depict a three domain wall growth morphology. In b) we
show the slope field composed of two right hand propagating domain
walls and a single propagating left hand domain wall. In a) we
show the corresponding moving facets in the height profile. Note
that the charges of the domain walls add up to zero implying a
flat interface at the edges, however, allowing for an offset}
\label{fig4}
\end{figure}

The interpretation of the time dependent growth morphology is
straightforward. For a dilute gas of growth modes the velocities
adjust to constant values after a transient period and the growth
modes move ballistically. Moreover, superimposed on the growth
modes is a gas of diffusive modes following from a linear analysis
of the equation of motion about the domain wall solutions, see
Ref. \cite{Fogedby03b}. In Fig.~\ref{fig4} we have depicted a
three domain wall growth configuration composed of interconnected
propagating domain walls, two right hand domain walls and one left
hand domain wall. We also show the resulting morphology in the
height field corresponding to moving steps or facets.

\begin{figure}
\includegraphics[width=1.0\hsize]{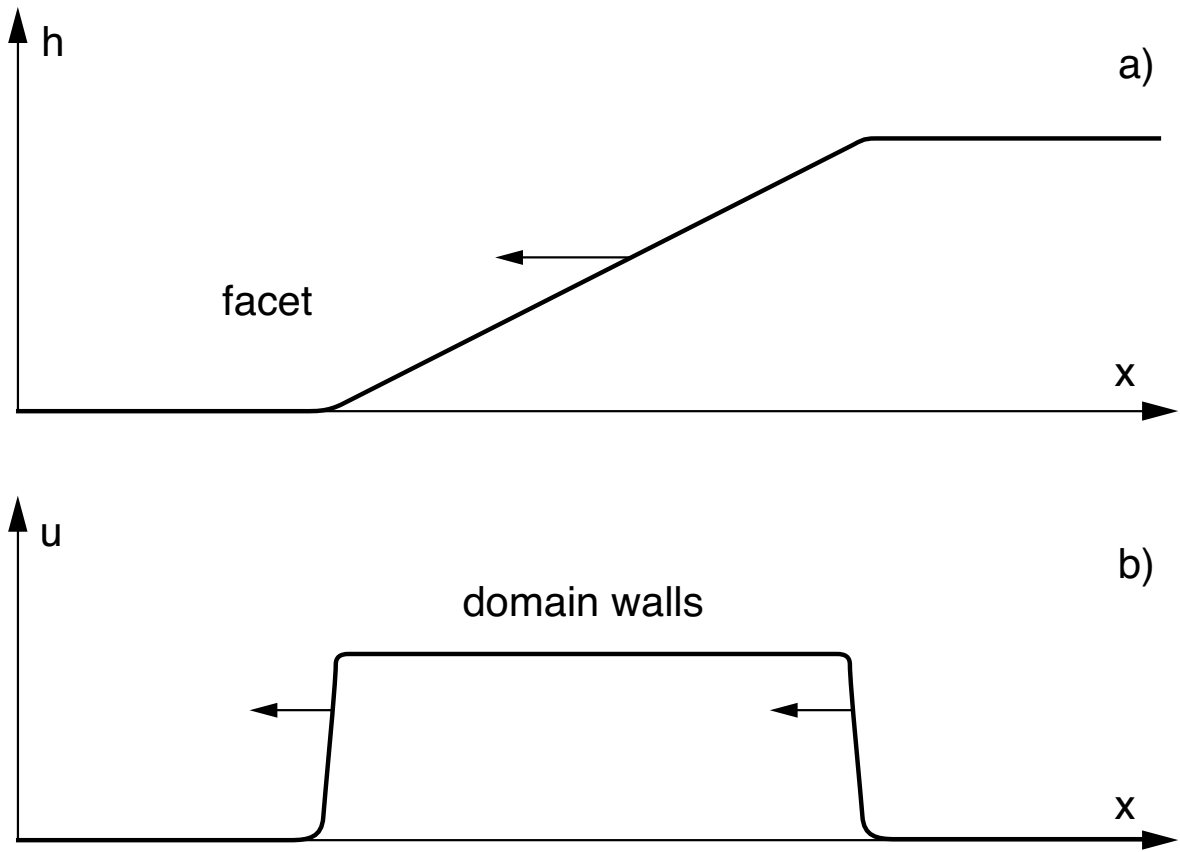}
\caption{In b) we show a co-moving two-domain wall configuration
in the slope $u$. This pair mode corresponds to a moving step or
facet in the height field $h$ depicted in a). The mode carries a
finite action associated with the left hand domain wall.}
\label{fig5}
\end{figure}

In order to make contact with the stochastic interpretation we
prepare the interface in a specific initial state $h(x,0)$
characterized by a gas of growth modes plus diffusive background.
By also assigning an appropriate noise field $p(x,0)$
corresponding to the nucleation of growth modes this configuration
propagates ballistically forward in time to a specific finite
configuration $h(x,T)$. Only the left hand domain walls
corresponding to negative charges carry an action. For a dilute
domain wall gas, ignoring the diffusive contribution, this action
is additive, i.e.,
\begin{eqnarray}
S=\frac{8\nu^2T}{3k_0^2}\sum_{k_i<0}|k_i|^3, \label{actsol}
\end{eqnarray}
yielding the transition probability
\begin{eqnarray}
P(h,T)\propto\exp\left[-\frac{S(h,T)}{\Delta}\right].\label{dissol}
\end{eqnarray}
For illustration consider the two-domain wall configuration
depicted in Fig.~\ref{fig5}. This pair mode has the form
\begin{eqnarray}
u(x,t)=\frac{k}{k_0}[u_+(x-vt-x_1)+u_-(x-vt-x_2)],
\label{pair}
\end{eqnarray}
and moves according to the domain wall matching condition following
from the Galilean symmetry with the velocity $v=-\lambda k/k_0$.
Since the pair mode in the slope $u$ corresponds to a moving step in
$h$ the propagation across the system either subject to periodic or
bouncing boundary condition corresponds to adding a layer to the
interface; the mode thus corresponds to a specific growth situation.
The mode moves ballistically with an action given by Eq.
(\ref{actdw}) carried by the left hand domain wall; note that the
right hand domain wall partner carries zero action. In time $T$ the
mode moves the distance $L=vT$ and we obtain the transition
probability
\begin{eqnarray}
P(L,T)\propto\exp
\left[-\frac{4\nu}{3\lambda^2\Delta}\frac{L^3}{T^2}\right].
\label{dispair}
\end{eqnarray}
We conclude that the step in $h$ performs a random walk with mean
square displacement
\begin{eqnarray}
<L^2>\propto(\lambda^2\Delta/\nu)^{2/3} T^{2/z}, \label{msqd}
\end{eqnarray}
characterized by the dynamical exponent $z=3/2$. This result is in
accordance with established scaling results for the KPZ equation
in 1D, see e.g. Ref. \cite{Halpin95}. The facet in the height
field corresponding to the pair growth mode thus performs
superdiffusion \cite{Feder88}.
\section{\label{mam}The Minimum Action Method}
In this section we discuss the basis for the minimum action method
characterized by the Freidlin-Wentzel action and the connection to
equivalent formulations in non equilibrium physics.
\subsection{Freidlin Wentzel scheme}
The point of departure for the Freidlin-Wentzel (FW) scheme is a
generic Langevin equation for a set of stochastic variables,
$\{x_n\}$, driven by additive white Gaussian noise
\begin{eqnarray}
\frac{dx_n}{dt}=-F_n(\{x_m\})+\eta_n(t), \label{lan2}
\end{eqnarray}
where the noise is distributed according to
\begin{eqnarray}
P(\{\eta_n)\},T)\propto\exp\left[-\frac{1}{2\Delta}\int_0^T dt
\sum_n\eta_n(t)^2\right]. \label{noise2}
\end{eqnarray}
A heuristic derivation of the Freidlin-Wentzel action, the basis
for the minimum action method, follows in the weak noise
limit by simply replacing the noise $\eta_n$ in Eq. (\ref{noise2})
by $dx_n/dt+F_n$ yielding
\begin{eqnarray}
P(\{x_n\},T)\propto\exp\left[-\frac{1}{2\Delta}\int_0^T
dt\sum_n\left(\frac{dx_n}{dt}+F_n\right)^2\right].~\label{dis2}
\end{eqnarray}
Expressing $P(\{x_n\},T)$ in the WKB form
\begin{eqnarray}
P(\{x_n\},T)\propto\exp\left[-\frac{S_{\text{FW}}}{\Delta}\right],
\label{dis22}
\end{eqnarray}
we readily identify the Freidlin-Wentzel action
\begin{eqnarray}
S_{\text{FW}}= \frac{1}{2}\int_0^T
dt\sum_n\left[\frac{dx_n}{dt}+F_n\right]^2; \label{fw}
\end{eqnarray}
for rigorous details see Refs. \cite{Freidlin98,E04}.

The minimum action method then corresponds to minimizing the
action $S_{\text{FW}}$ subject to an initial condition
$\{x_n(0)\}$, a final condition $\{x_n(T)\}$, and a given
transition time $T$. The method thus identifies the minimum action
path in the ``action landscape''. The method works both for
gradient systems where $F_n$ can be derived from a free energy,
\begin{eqnarray}
F_n=\nabla_n\Phi, \label{free}
\end{eqnarray}
including e.g. the GL case and non-gradient systems like the KPZ
equation.
\subsection{Martin Siggia Rose scheme}
The Martin-Siggia-Rose (MSR) scheme
\cite{Martin73,deDominicis75,deDominicis76,
Baussch76,Janssen76,deDominicis78} also takes as its starting
point the Langevin equation (\ref{lan2}). For simplicity we
consider, however, only a single stochastic variable $x(t)$. For
the transition probability $P(x,T)$ we have by definition
\begin{eqnarray}
P(x,T)=\langle\delta(x-x(T))\rangle_\eta, \label{dis3}
\end{eqnarray}
where we average over the noise $\eta$ driving the Langevin
equation. Incorporating the Langevin equation determining the
evolution of $x(t)$ as a delta function constraint, averaging over
the noise $\eta$ according to Eq. (\ref{noise2}), noting that the
change of variable from  $dx/dt$ to $x$ yields the Jacobian
$J=\exp[(1/2)\int dt dF/dx]$, and finally setting $p\rightarrow
p/\Delta$ we obtain the functional phase space integral
\cite{Zinn-Justin89}
\begin{eqnarray}
P(x,T)\propto\int\prod_tdxdp~\delta(x-x(T))
\exp\left[-i\frac{S_{\text{MSR}}}{\Delta}\right],~~ \label{path1}
\end{eqnarray}
where the MSR action is given by
\begin{eqnarray}
S_{\text{MSR}}=\int dt\left[p\frac{dx}{dt}-H\right], \label{amsr}
\end{eqnarray}
with MSR Hamiltonian
\begin{eqnarray}
H_{\text{MSR}}=-\frac{i}{2}p^2-pF+\frac{i\Delta}{2}\frac{dF}{dx}.
\label{hmsr}
\end{eqnarray}
Since $p$ appears quadratically it can be eliminated by a Gaussian
integration and we arrive at the configuration space path integral
\begin{eqnarray}
P(x,T)\propto\int\prod_tdx\delta(x-x(T))\exp\left[-\frac{S}{\Delta}\right],
\label{dis4}
\end{eqnarray}
with action
\begin{eqnarray}
S=\frac{1}{2}\int
dt\left[\left(\frac{dx}{dt}+F\right)^2-\Delta\frac{dF}{dx}\right].
\label{afw}
\end{eqnarray}
We note that this form holds for arbitrary noise strength. The
path integral is a formal solution of the Fokker-Planck equation.
In the aymptotic weak noise limit $\Delta\rightarrow 0$ only the
saddle point in the path integral contributes. Ignoring the
Jacobian contribution $\Delta dF/dx$ we recover the FW result in
Eq. (\ref{dis2}) in the case of one variable.
\subsection{Quantum analogue and phase space method}
\begin{figure}
\includegraphics[width=1.0\hsize]{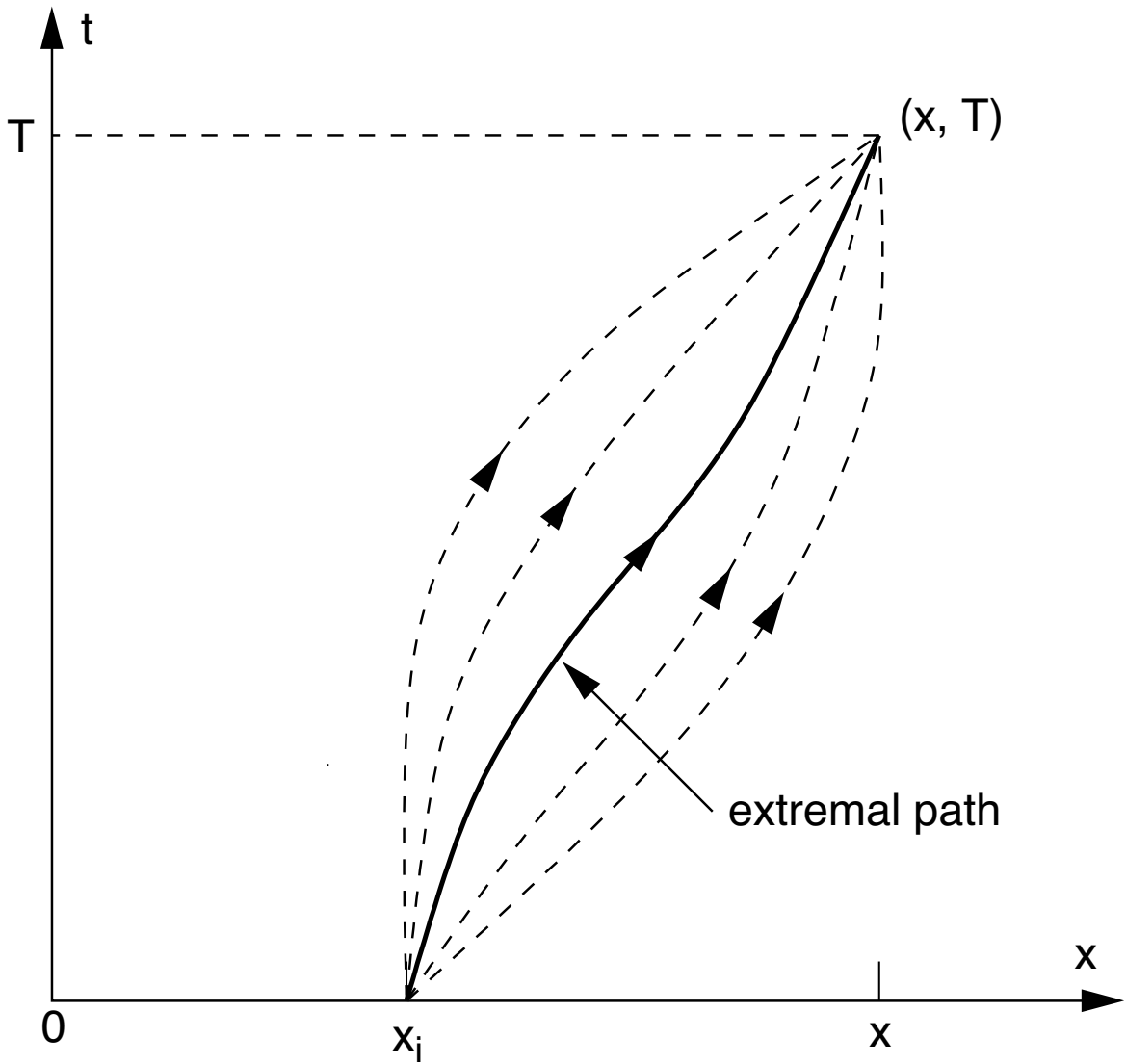}
\caption{We show paths (dashed) from an initial configuration
$x_i$ at time $t=0$ to a final configuration $x$ at time $t=T$
contributing to the path integral. We also depict the extremal
path (solid) dominating the path integral in the limit of weak
noise.} \label{fig6}
\end{figure}

Contact with the Fokker-Planck equation (\ref{fp}) is easily
achieved by noting that Eq. (\ref{path1}) has the form of a
Feynmann path integral with Planck constant $\Delta$
\cite{Zinn-Justin89,Feynman65,Das93}. Introducing the momentum
operator $\hat p=-i\Delta d/dx$ the quantum Hamiltonian operator
takes the form
\begin{eqnarray}
\hat H = \frac{i}{2}\Delta^2\frac{d^2}{dx^2}+
\left(i\Delta\frac{dF}{dx}\right)_{\text{order}}
+\frac{i\Delta}{2}\frac{dF}{dx}, \label{qham}
\end{eqnarray}
where the ordering in the term $(i\Delta dF/dx)_{\text{order}}$
remains to be fixed. Choosing the symmetrical Weyl ordering
$(dF/dx)_{\text{order}}=(1/2)(Fd/dx+dF/dx)$ the Schr\"odinger
equation associated with $\hat H$,
\begin{eqnarray}
i\Delta\frac{\partial P}{\partial t}=\hat H P, \label{schr}
\end{eqnarray}
then reduces to the Fokker-Planck equation (\ref{fp}). Finally,
formally rotating $p$, $p\rightarrow ip$, we obtain a real path
integral representation for $P$,
\begin{eqnarray}
P(x,T)\propto\int\prod_t
dxdp\delta(x-x(T))\exp\left[-\frac{S}{\Delta}\right], \label{path2}
\end{eqnarray}
with action
\begin{eqnarray}
S=\int dt \left[p\frac{dx}{dt}-H\right] \label{act3}
\end{eqnarray}
and Hamiltonian
\begin{eqnarray}
H=\frac{1}{2}p^2-pF+\frac{\Delta}{2}\frac{dF}{dx}. \label{ham3}
\end{eqnarray}
In the weak noise limit $\Delta\rightarrow 0$ the Jacobian
contribution in Eq. (\ref{ham3}) can be ignored and only the
saddle point in Eq. (\ref{path2}) contributes, corresponding to a
principle of least action $\delta S=0$. In this way we recover the
results in Section II C. We note that the canonical phase space
method is completely equivalent to the Freidlin-Wentzel scheme for
the extremal orbits. In fact inserting Eq. (\ref{lan2}) for one
variable, $dx/dt=-F+p$ in Eq. (\ref{act2}) we obtain $S=(1/2)\int
dt p^2$ which is the Freidlin-Wentzel action. The advantage of the
phase space method is the introduction of the canonically
conjugate momentum $p$, representing the noise, as an additional
variable. This allows for a phase space representation of the
numerical results obtained by a numerical optimization of the
Freidlin-Wentzel action. In Fig.~\ref{fig6} we have in a $xt$ plot
depicted the paths in configuration space from an initial
configuration $x_i$ at time $t=0$ to a final configuration $x$ at
time $t=T$. We have also shown the extremal path which dominates
the path integral in the limit $\Delta\rightarrow 0$.
\section{\label{num}Minimum action method for the KPZ equation}
In this section we apply the minimum action method to the KPZ
equation and set up the numerical scheme. For the KPZ equation the
FW action has the form
\begin{eqnarray}
S=\frac{1}{2}\int{\bf dr}dt \left(\frac{\partial h}{\partial t}
-\nu{\bm\nabla}^2h -\frac{\lambda}{2}({\bm\nabla} h)^2+F\right)^2.
\label{actkpz}
\end{eqnarray}
In order to find the optimal switching path from an initial
configuration $h_{\text{init}}({\bf r})$ at time $t=0$ to a final
configuration $h_{\text{fin}}({\bf r})$ at time $T$ we minimize
the action (\ref{actkpz}) subject to the constraints:
\begin{equation}
h({\bf r},0)=h_{\text{init}}({\bf r}),\ \
h({\bf r},T)=h_{\text{fin}}({\bf r}).
\end{equation}
We first discretize the action functional using finite
differences, then minimize the discretized action functional using
the limited memory Broyden-Fletcher-Goldfarb-Shanno (BFGS) method.
The BFGS is an efficient quasi-Newton method for large scale
optiminization problems \cite{Nocedal99}. It is an iterative
method; at each iteration, it only requires the input of the
action $S$ and the associated gradient $\delta S/\delta h({\bf
r},t)$. The minimization is constrained by appropriate Dirichlet
boundary conditions in space, $h({\bf r},t)=h_{\cal B}({\bf r})$
for ${\bf r}$ on boundary ${\cal B}$, and initial and final
boundary conditions in time, $h({\bf r},0)=h_{\text{init}}({\bf
r})$ and $h({\bf r},T)=h_{\text{fin}}({\bf r})$

In the following we consider the 1D case. We confine the system to
a 1D interval of size $L$ and the switching path to the time interval
$T$. In the space-time domain $\left[0,L\right]\times \left[0,T\right]$
we introduce a mesh with sizes
$\Delta x=L/I$ and $\Delta t=T/J$ and define the grid points
$(x_i,t_j)$
\begin{eqnarray}
&&x_i=i\Delta x~~~~i=0,\cdots I, \label{disx}
\\
&&t_j=j\Delta t~~~~j=0,\cdots J. \label{dist}
\end{eqnarray}
The numerical approximation to $h(x_i,t_j)$ is denoted by $H_{ij}$.
In order to simplify the expression we introduce the momentum or
the noise field
\begin{eqnarray}
p(x,t)=\frac{\partial h}{\partial t}
-\nu\frac{\partial^2h}{\partial x^2}- \frac{\lambda}{2}
\left(\frac{\partial h}{\partial x}\right)^2+F, \label{force}
\end{eqnarray}
and express the action in the form
\begin{eqnarray}
S(h)=\frac{1}{2}\int_0^Tdt\int_0^Ldx p^2(x,t). \label{act4}
\end{eqnarray}
Using the trapezoidal rule to discretize the integral in space and
the midpoint rule to compute the temporal integral we obtain
\begin{eqnarray}
S(H)= \frac{1}{2}\Delta x\Delta t
\sum_{i=1}^{I-1}\sum_{j=1}^{J}P_{ij}^2, \label{actnum}
\end{eqnarray}
where the discretized version of the noise field is
\begin{eqnarray}
&&P_{ij} = \frac{ H_{ij}-H_{i,j-1}}{\Delta t} +F \nonumber
\\
&&-\nu\frac{H_{i+1,j}+H_{i-1,j}-2H_{ij}}{2(\Delta x)^2}\nonumber
\\
&&-\nu\frac{H_{i+1,j-1}+H_{i-1,j-1}-2H_{i,j-1}}{2(\Delta x)^2}\nonumber
\\
&&-\frac{\lambda}{2}
\frac{(H_{i+1,j}-H_{i-1,j}+H_{i+1,j-1}-H_{i-1,j-1})^2}{16(\Delta
x)^2}.~~~~ \label{forcenum}
\end{eqnarray}
For the discretized boundary condition we have
\begin{eqnarray}
&&H_{0j}=H_1,~~H_{Ij}=H_2~~\text{for}~~j=0,\cdots J~~~~~~~~
\\
&&H_{i0}=h_{\text{init}}(x_i),~~
H_{iJ}=h_{\text{fin}}(x_i)~~\text{for}~~i=0,\cdots I,~~~~~~~
\label{boundary}
\end{eqnarray}
where $H_1$ and $H_2$ denotes the boundary values. For an offset
in the height profile we have $H_1\neq H_2$. The BFGS method also
requires the gradient of the action, whose discrete version is
given by
\begin{eqnarray}
&&\frac{\partial S}{\partial H_{ij}} =  \Delta x\Delta t
\Biggl(\frac{ P_{ij}-P_{i,j+1}}{\Delta t} \nonumber
\\
&&-\nu\frac{P_{i+1,j}+P_{i-1,j}-2P_{ij}}{2(\Delta x)^2}\nonumber
\\
&&-\nu\frac{P_{i+1,j+1}+P_{i-1,j+1}-2P_{i,j+1}}{2(\Delta
x)^2}\nonumber
\\
&&-\frac{\lambda}{2}\frac{(H_{ij}-H_{i-2,j}+
H_{i,j+1}-H_{i-2,j+1})P_{i-1,j+1}}{8(\Delta x)^2} \nonumber
\\
&&+\frac{\lambda}{2}\frac{(H_{i+2,j}-H_{ij}+
H_{i+2,j+1}-H_{i,j+1})P_{i+1,j+1}}{8(\Delta x)^2} \nonumber
\\
&&-\frac{\lambda}{2}\frac{(H_{i,j-1}-H_{i-2,j-1}+
H_{ij}-H_{i-2,j})P_{i-1,j}}{8(\Delta x)^2} \nonumber
\\
&&+\frac{\lambda}{2}\frac{(H_{i+2,j-1}-H_{i,j-1}+
H_{i+2,j}-H_{ij})P_{i+1,j}}{8(\Delta x)^2}\Biggr). \nonumber
\\
&&
 \label{numforce}
\end{eqnarray}
The numerical optimization is set up by choosing an initial
pathway interpolating between the initial and final configurations
$h_{\text{init}}$ and $h_{\text{fin}}$ subject to the chosen
boundary conditions. Provided that the initial pathway lies in the
domain of the appropriate minimum of $S$ the BFGS method then
through successive steps finds the minimum action and outputs
the weak noise pathway from $h_{\text{init}}(x)$ to
$h_{\text{fin}}(x)$ in a given transition time $T$.
\section{\label{numres}Numerical results in 1D}
In this section we discuss various switching scenarios for the KPZ
equation in 1D. As parameter values we choose for the viscosity
$\nu=1$ and for the non-linear growth parameter $\lambda=2$. These
values yield the inverse length scale $k_0=1$. The parameter $k$
is then given by $k=\sqrt F$ where $F$ is the imposed drift. We,
moreover, consider a system of size $L=1$.

We consider the switching scenario in 1D from an initial state
$h(x,0)=-h_0$ to a final state $h(x,T)=h_0$. This transition
corresponds to adding a layer of thickness $2h_0$ to the
interface. At the boundaries $x=0$ and $x=L$ we set $h=0$, i.e.,
$H_1=H_2=0$. In order to match the initial profile
$h_{\text{init}}(x)=-h_0$ to the boundary condition we use the
cusp solutions in Eq. (\ref{sol2}), $h_\pm(x)=\pm(1/k_0)\ln |\cosh
kx|$ and set $h(x,0)=h_L(x)+h_R(x)$, where
\begin{eqnarray}
&& h_L(x)=-\frac{1}{k_0}\ln \left|\frac{\cosh k(x-x_1)}{\cosh
k(x-x_1-\delta)}\right|,   
\label{leftinit}
\\
&& h_R(x)=~~\frac{1}{k_0}\ln \left|\frac{\cosh k(x-x_2)}{\cosh
k(x-x_2-\delta)}\right|.   
\label{rightinit}
\end{eqnarray}
Setting $x_1\sim 0$ and $x_2\sim L-\delta$  and choosing
$\delta=h_0k_0/2k$ the initial profile satisfies the boundary
conditions and approach the interface value $-h_0$ in the bulk;
note that the slope of the steps is given by $1/k$. In our
simulation we choose $x_1=0.1$, $x_2=0.8$, $\delta =0.1$.
Likewise, the final configuration at time $t=T$ is given by
$h(x,T) =-h_L(x)-h_R(x)$. To ensure a steep step corresponding to
a short healing length we choose the drift $F=625$ corresponding
to $k=25$. With this choice $h_0=2k\delta/k_0=5$. For the initial
path, we use the linear interpolation between $h(x,0)$ and
$h(x,T)$: $h(x,t)=(1-t/T)h(x,0)+(t/T)h(x,T)$. Finally, we have
chosen a $200\times 200$ set of $x t$ grid points.

\begin{figure}
\includegraphics[width=0.8\hsize]{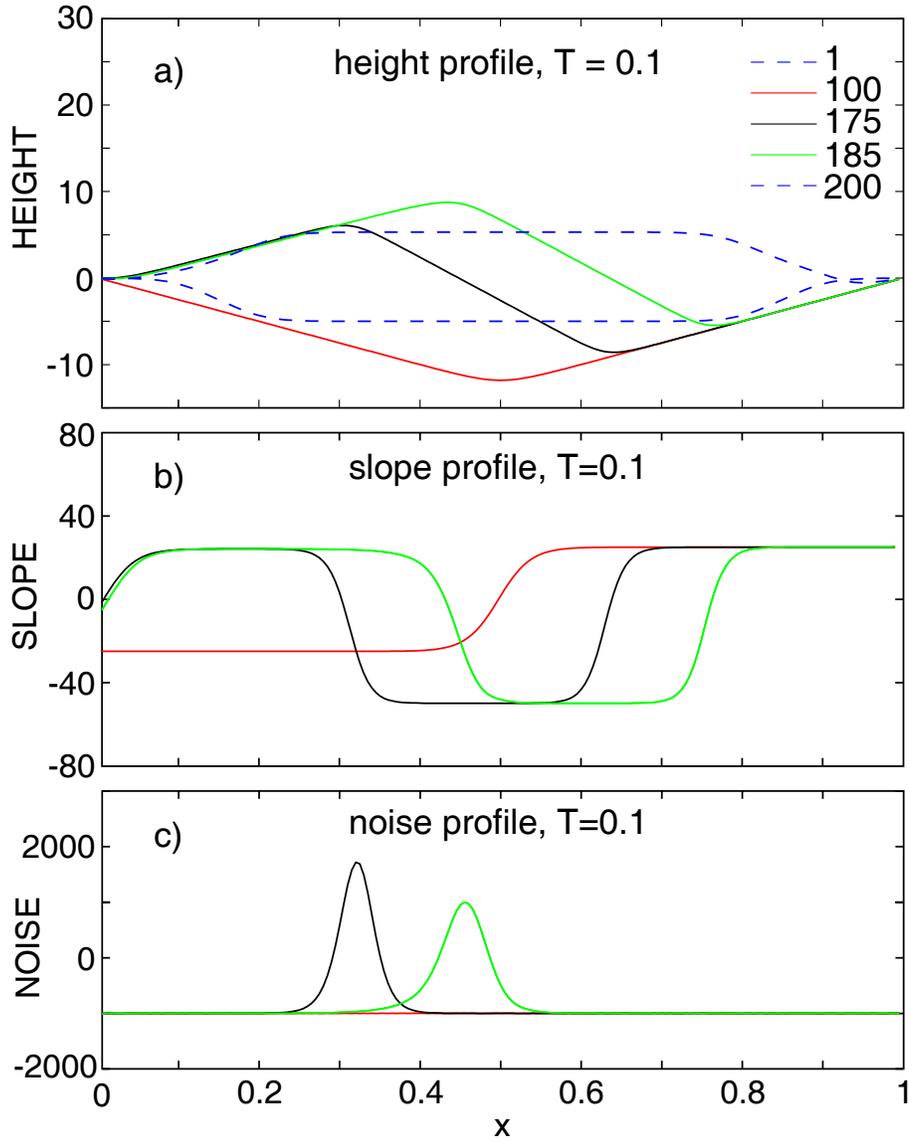}
\caption{We depict the transition scenario for transition time
$T=0.1$. In a) we show the waiting time configuration and the
propagating step in $h$, in b) the quasi static right hand domain
wall and the propagating domain wall pair in $u$, and in c) the
corresponding noise field associated with the propagating left
hand domain wall (arbitrary units).} \label{fig7}
\end{figure}

In Figs.~\ref{fig7}- \ref{fig10}  we show switching scenarios for
the transition times $T=0.1$, $T=0.03$, $T=0.01$, and $T=0.001$.
We depict both the height profiles $h(x,t)$, the slope profiles
$u(x,t)$, and the associated noise profiles $p(x,t)$. In
Figs.~\ref{fig11}-\ref{fig14} we depict the associated squared
noise field or action density in a space-time plot.

The height profiles presented for the initial and final
configurations $h_{\text{init}}$ and $h_{\text{fin}}$ and for some
characteristic intermediate times show that the transition or
switch in time $T$ is effectuated by the ballistic propagation of
steps or facets across the system. The corresponding slope
profiles demonstrate that the steps can be interpreted in terms of
a gas of domain walls with opposite parity, i.e., right hand and
left hand domain walls. The motion of a single step in $h$ is thus
associated with a pair of co-moving domain walls in $u$ moving
across the system. The dependent noise field $p$ is associated
with the nucleation of domain walls. Since the right-hand domain
wall is a solution of the deterministic Burgers equation it
carries no dynamical attributes and the associated noise field
vanishes, unlike the ``noise-induced'' left-hand domain wall which
is associated with a noise field and carries an action. In Table I
we show the actions associated with the transitions and in
Figs.~\ref{fig15} the action as a function of the transition time
for the various scenarios.
\begin{table}
\begin{tabular}{|c|c|}
  \hline
  ~~Transition time $T$~~ & ~~Switching action $S$~~ \\
  \hline
  0.100 &  $2.57\times 10^3$   \\
  0.030 &  $2.56\times 10^3$   \\
  0.010 &  $3.12\times 10^3$   \\
  0.001 &  $1.95\times 10^4$  \\
    \hline
\end{tabular}
\caption{The switching actions $S(T)$ associated with the
transition times $T =0.100, 0.030, 0.010, 0.001$.\label{table1:}}
\end{table}
\section{\label{disc}Discussion and interpretation}
In this section we interpret the numerical results in 1D using the
weak noise canonical phase method. As discussed in Sec. \ref{mam}
the phase space method is completely equivalent to the minimum
action method.
\subsection{Waiting time transition for $T=0.1$}
In terms of the switching dynamics $T=0.1$ corresponds to a
long-time transition. In Fig.~\ref{fig7} we show snapshots of $h$,
$u$, and $p$ at times $t=~$0.0, 0.05, 0.0875, 0.0925, 0.1; in
Fig.~\ref{fig11} we depict the squared noise field or space-time
action density. In the initial stage of the transition, from $t=0$
to about $t=0.075$, the constant height field makes a transition to
a trough (convex cusp) compatible with the boundary conditions
$h=0$. This configuration corresponds to a static right hand
domain wall in the slope $u$. After a long waiting time in this
configuration (until about $t=0.075$)
domain walls in $u$ nucleate at the boundaries and a pair of
domain walls then move across the system from left to right. In the
height field this mode corresponds to the motion of a facet or
step. The trough in $h$ is filled in and eventually at time $T$
the final configuration $h_{\text{fin}}$ is reached. The noise
field associated with the waiting time configuration vanishes
since it corresponds to a right hand domain wall. However, in the
final stage of the transition the noise field develops
corresponding to the nucleation of the left hand domain wall.

This switching scenario is in accordance with the phase space
interpretation generically represented in Fig.~\ref{fig2}. For a
long time transition the orbit comes close to the saddle point
corresponding to $p=0$. In the slope field this implies a
configuration given by the right hand domain wall $u=(k/k_0)\tanh
kx$ yielding the cusp in Fig.~7a. After a long waiting time in the
vicinity of the saddle point the orbit eventually wanders off along
the stationary manifold towards the final configuration. This part
of the orbit in phase phase associated with a finite noise field
corresponds to the propagation of the step in $h$, associated with
the domain wall pair in $u$.

The action can also be estimated qualitatively. For a single
left-hand domain wall the action is given by Eq. (\ref{actdw}),
$S_{\text{dw}}= (8/3)\nu^2(k^3/k_0^2)T$. Inserting $\nu=1$,
$k_0=1$, and $k=25$ we obtain $S_{\text{dw}}=41667\times T$.
However, owing to the waiting time only the last $p\neq 0$ part of
the orbit contributes to $S_{\text{dw}}$. Estimating the effective
transition time to be $T\sim 0.05$ we obtain an action of order
$S_{\text{dw}}\sim 2000$ which should be compared with the
numerical value from Table I, $S_{\text{num}}=2567$. The
discrepancy can be accounted for by the finite nucleation action
at the boundaries and also the finite system size effect.
\subsection{Intermediate time transitions, $T=0.03$ and $T=0.01$}
\begin{figure}
\includegraphics[width=0.8\hsize]{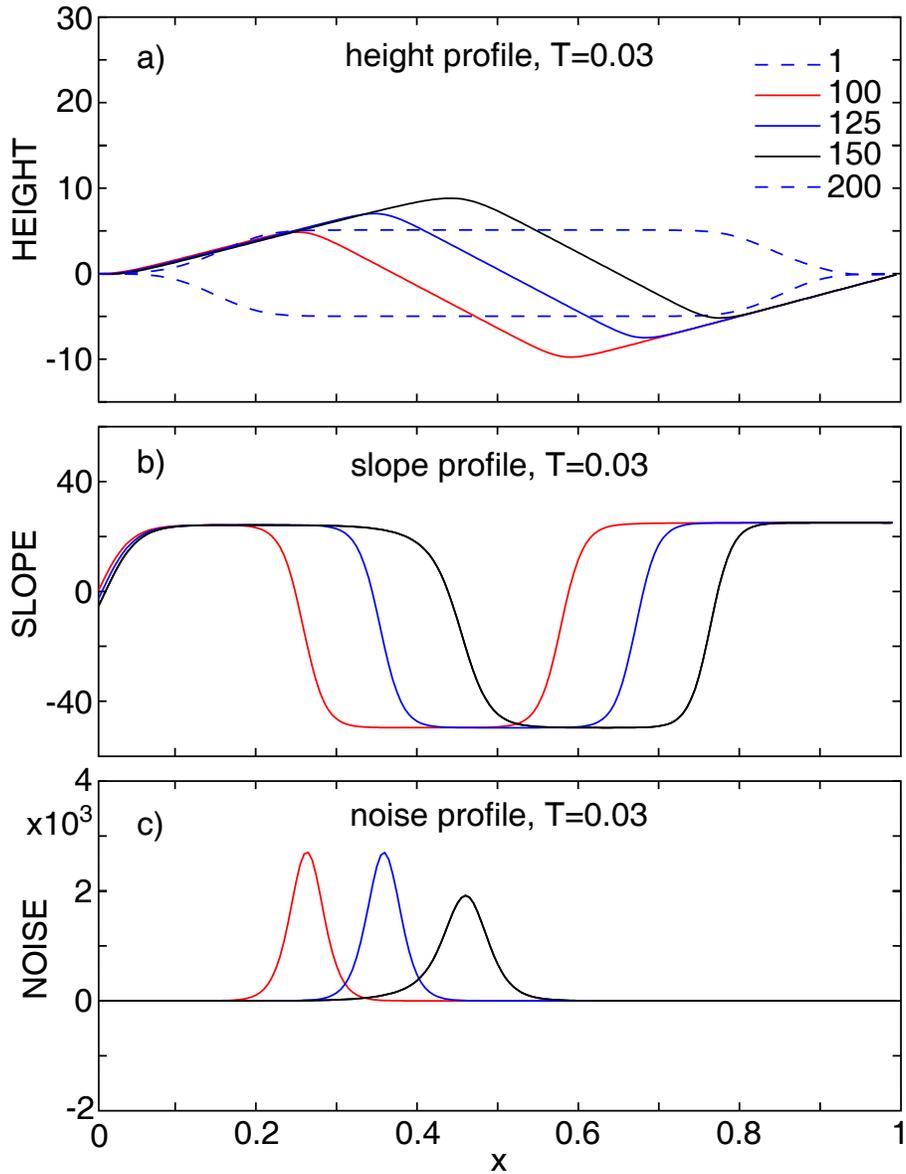}
\caption{We depict the transition scenario for transition time
$T=0.03$. In a) we show the propagating step in $h$, in b) the
propagating domain wall pair in $u$, and in c) the corresponding
noise field associated with the propagating left hand domain wall
(arbitrary units).} \label{fig8}
\end{figure}

\begin{figure}
\includegraphics[width=0.8\hsize]{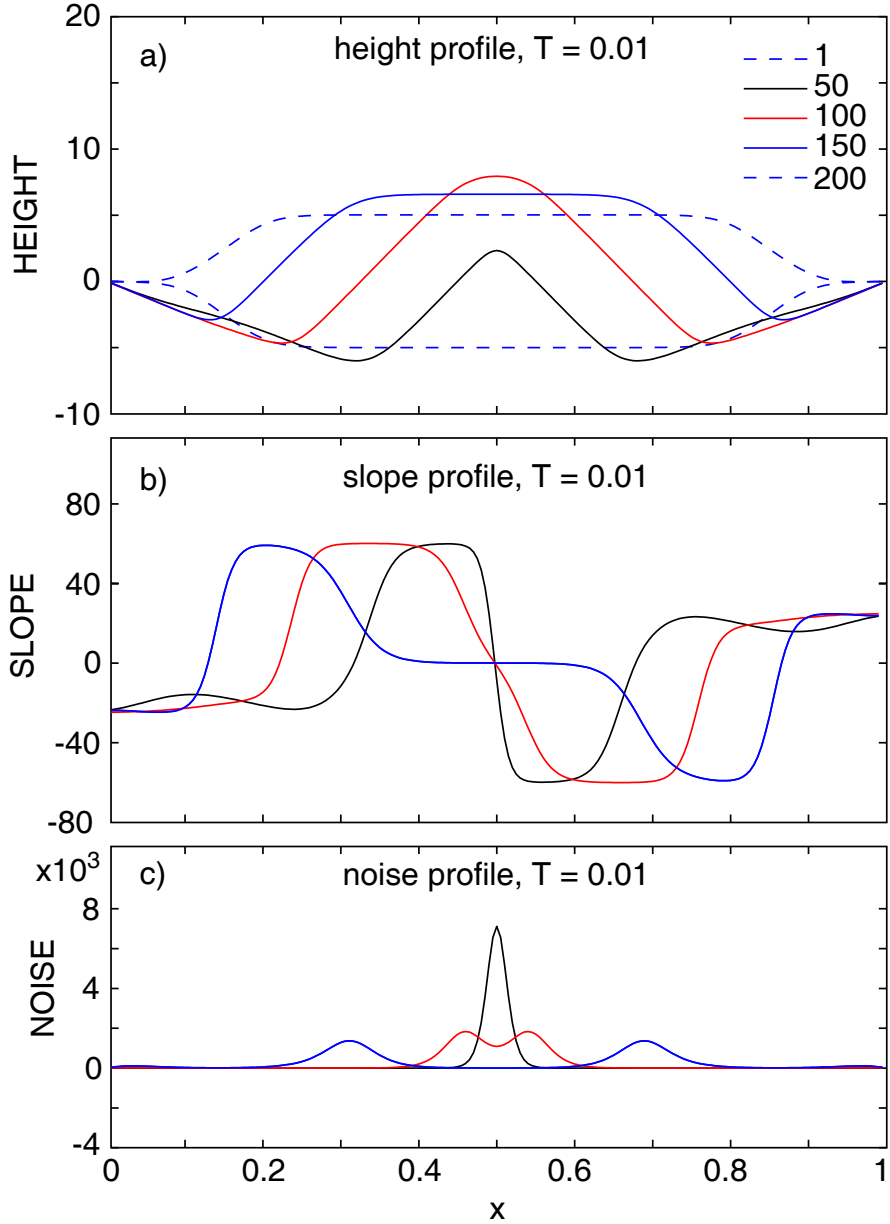}
\caption{We depict the transition scenario for transition time
$T=0.01$. In a) we show the emerging plateau in $h$, in b) the
left hand domain wall associated with the appearance of the peak
in $h$ and the propagating domain wall pairs emerging from the
center in $u$, and in c) the corresponding noise field associated
with the nucleation and subsequent propagation from the center
(arbitrary units).} \label{fig9}
\end{figure}

In Figs~\ref{fig8} and \ref{fig9} we have depicted switching
scenarios at transition times $T=0.03$ and $T=0.01$ for the
height, slope and noise. In Fig.~\ref{fig8} we show snapshots
along the pathway at times $t= 0.0, 0.015, 0.0188, 0.0225, 0.03$
and in Fig.~\ref{fig9} at times $t= 0.0, 0.0025, 0.005, 0.0075,
0.01$. In Figs.~\ref{fig12} and \ref{fig13} we depict the squared
noise field or space-time action density. Since the imposed
transition time is shorter compared to the previous case the
waiting time is shortened. The transition again is driven
by the nucleation and subsequent propagation of domain walls. In
the case $T=0.03$ domain walls in $u$ are nucleated at the edges
and the pair propagates across the system with a positive velocity
similar to the waiting time case. In the case $T=0.01$ the shorter
transition time favors the nucleation of a domain wall in $u$ at
the center. This domain wall subsequently breaks up into two pairs
of domain wall moving toward the edges. In the height profile this
scenario corresponds to the nucleation of a tip which subsequently
broadens to a plateau effectuating the transition.

This switching scenario is again heuristically in agreement with
the phase space interpretation in Fig.~\ref{fig2}. For an
intermediate time transition the orbit in phase space bends off
towards the stationary finite $p$ manifold at an earlier stage in
order to effectuate the transition in the shorter time interval
available.

The action based on Eq. (\ref{actdw}) is again of the same order
of magnitude as the numerical results listed in Table I. We note
that the shorter transition time requires a larger domain wall
velocity $v\propto k_i$, where $k_i$ is the charge of the
particular domain wall. Since the action scales with $k_i^3$ this
effect compensates in the action for the smaller $T$. For an
infinite system the imposed drift $F\propto k^2$ in the KPZ
equation is related to the domain wall charges $k_i$ by the
relationship $k=\sum_ik_i$. Due to finite size effects this
relation cannot be used directly in the present context. However,
we still conclude that the imposed $k$ does not fix the individual
charges. The domain wall amplitudes and thus velocities are
determined by the transition scenario.
\subsection{Short time transition, $T=0.001$}
\begin{figure}
\includegraphics[width=0.8\hsize]{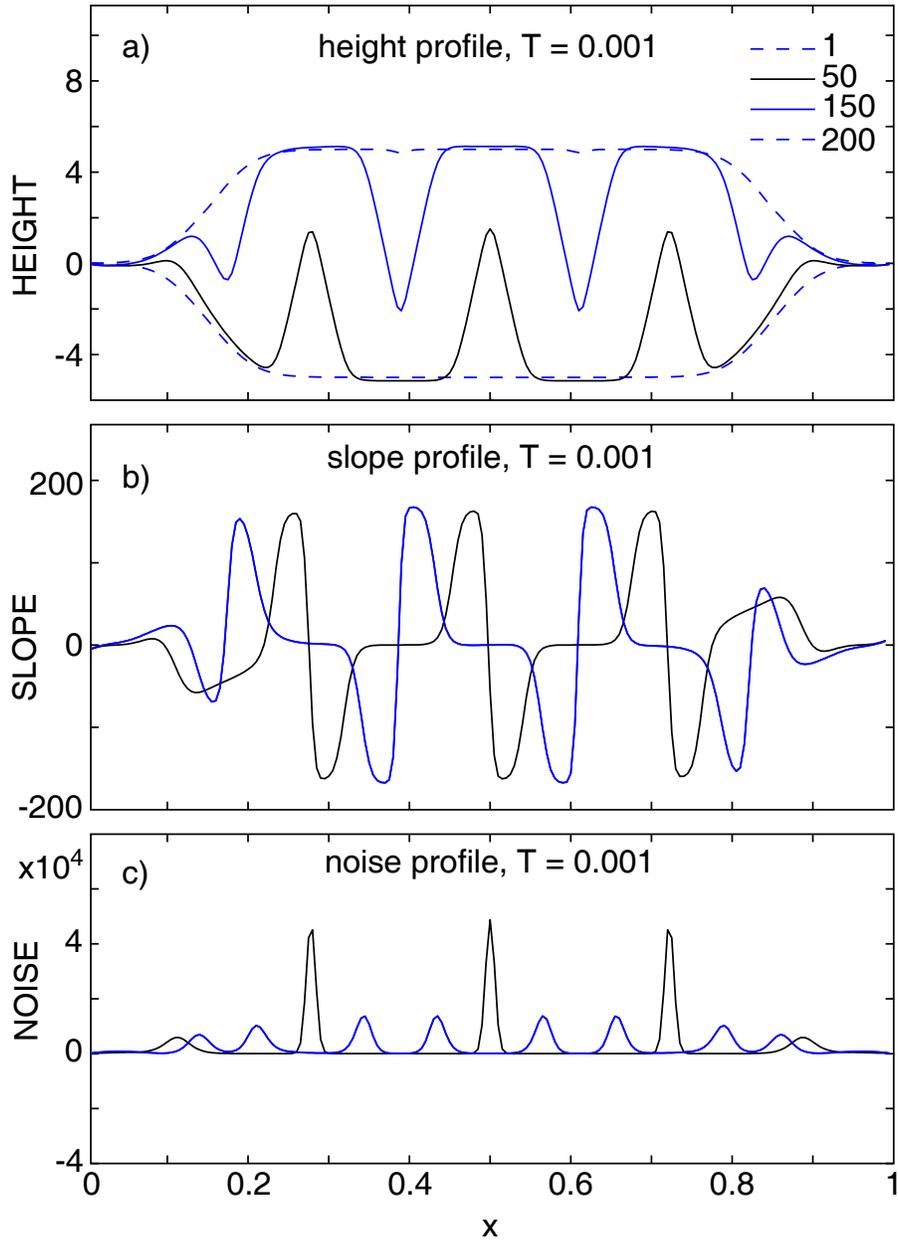}
\caption{We depict the transition scenario for transition time
$T=0.001$. In a) we show the propagation of the multiple steps or
facets in $h$, in b) the associated domain wall pairs in $u$, and
in c) the corresponding noise field associated with the nucleation
and subsequent propagation of domain walls (arbitrary units).}
\label{fig10}
\end{figure}

\begin{figure}
\includegraphics[width=0.8\hsize]{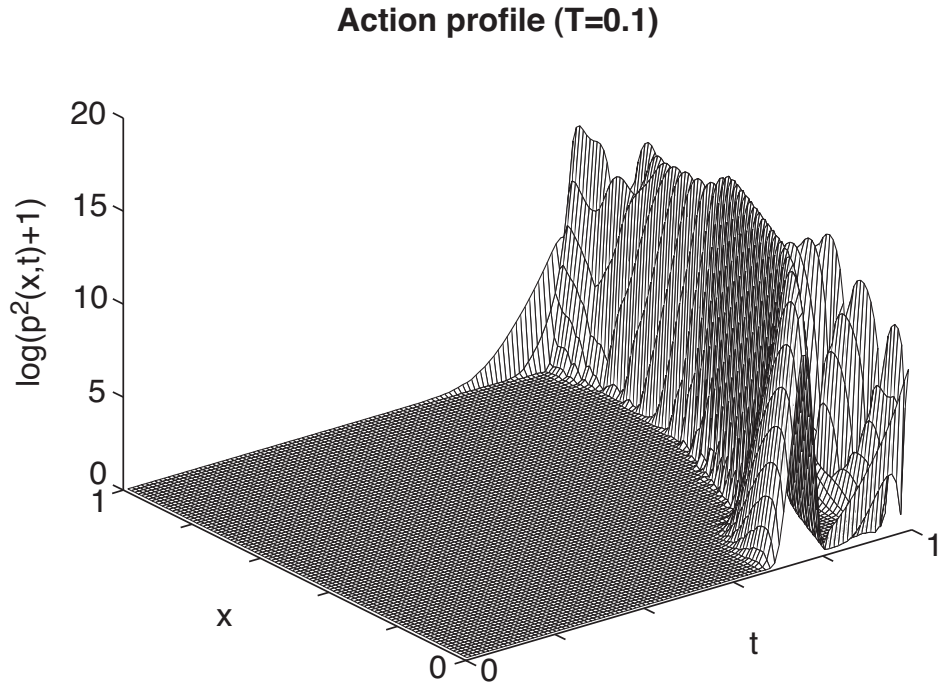}
\caption{We plot the squared noise field $p(x,t)^2$ or action
density as a function of $x$ and $t$ in the case $T=0.1$. The plot
show the waiting time aspects of the transition scenario
(arbitrary units).} \label{fig11}
\end{figure}

\begin{figure}
\includegraphics[width=0.8\hsize]{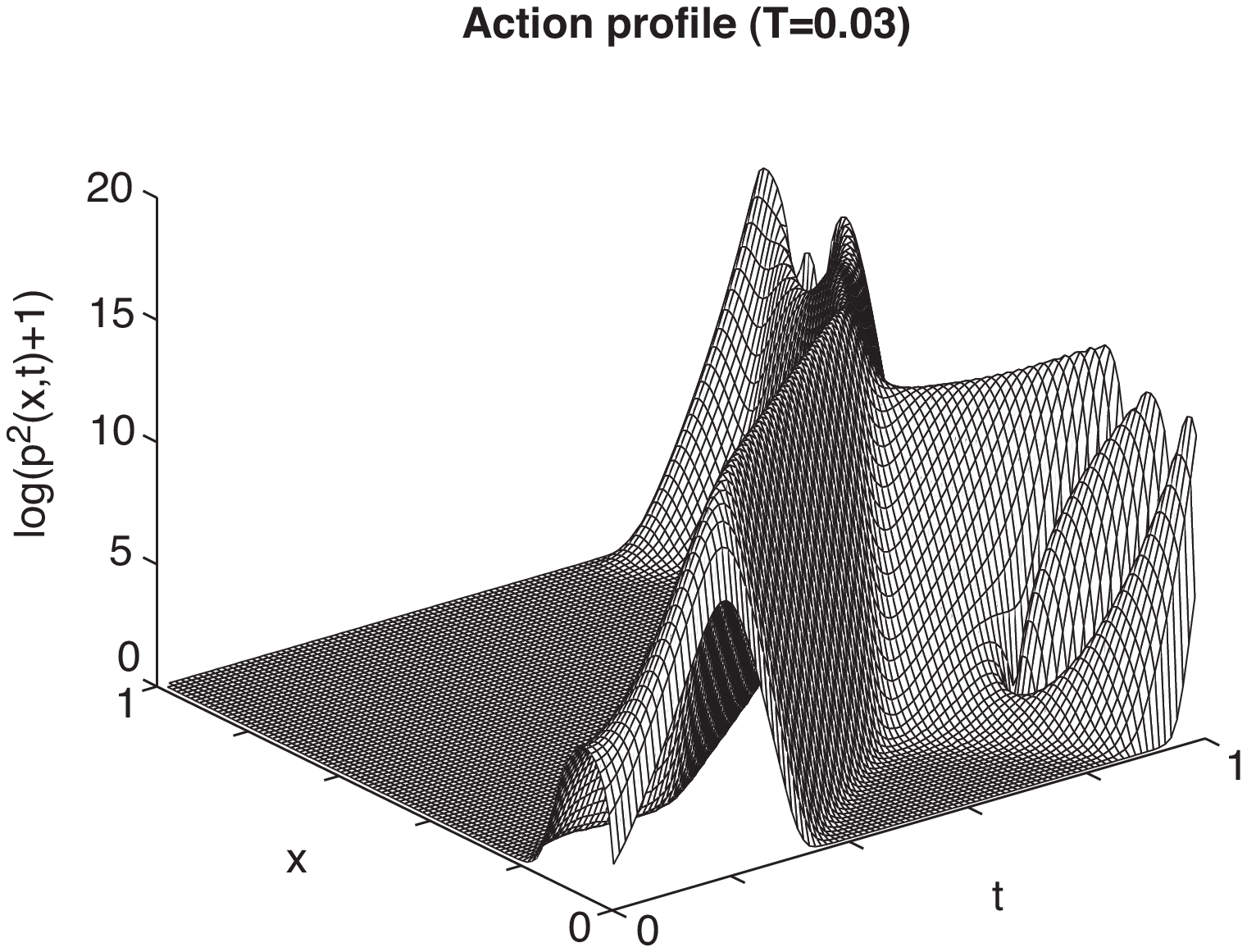}
\caption{We plot the squared noise field $p(x,t)^2$ or action
density as a function of $x$ and $t$ in the case
$T=0.03$(arbitrary units).} \label{fig12}
\end{figure}

\begin{figure}
\includegraphics[width=0.8\hsize]{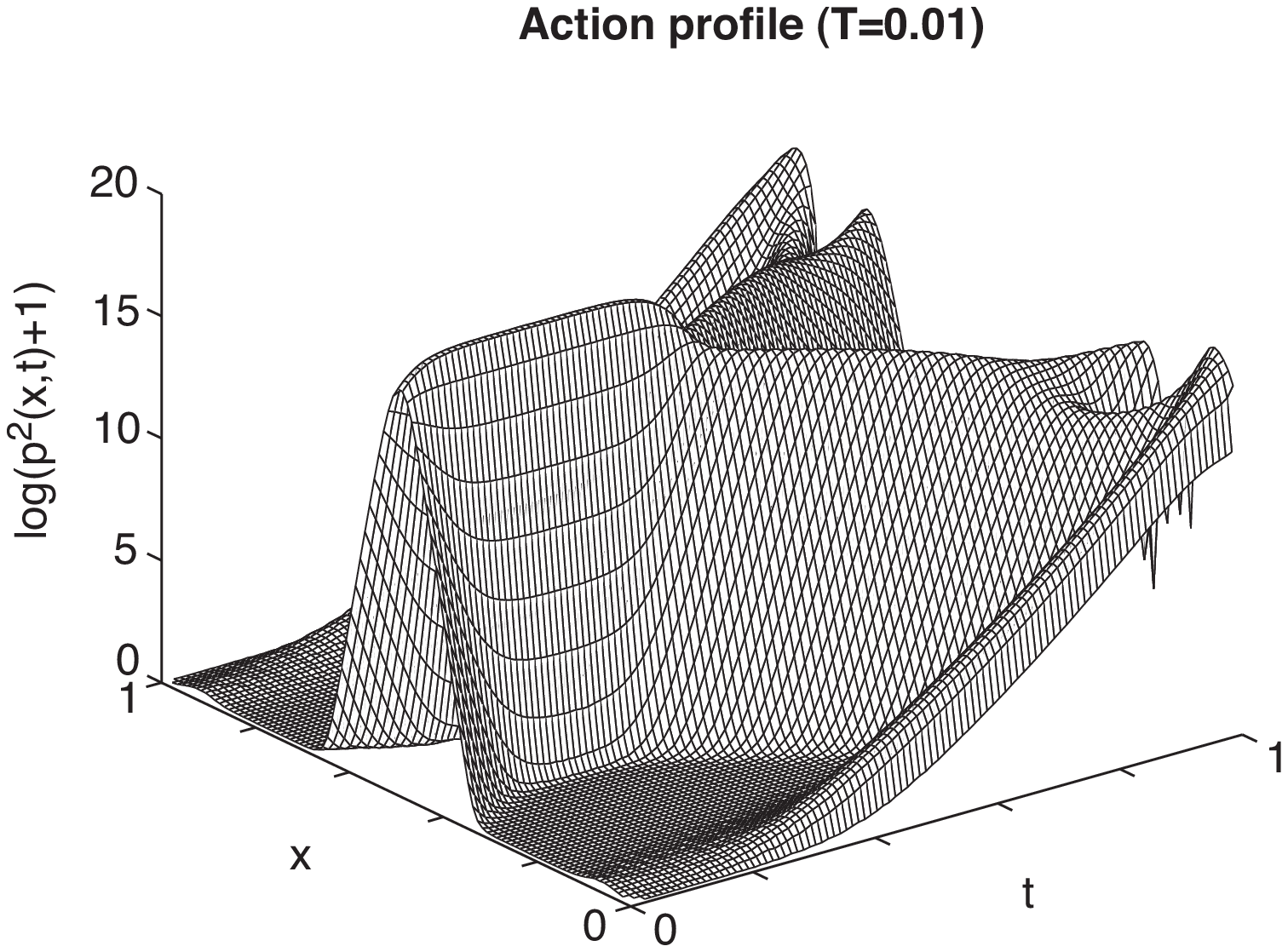}
\caption{We plot the squared noise field $p(x,t)^2$ or action
density as a function of $x$ and $t$ in the case $T=0.01$
(arbitrary units).} \label{fig13}
\end{figure}

\begin{figure}
\includegraphics[width=0.8\hsize]{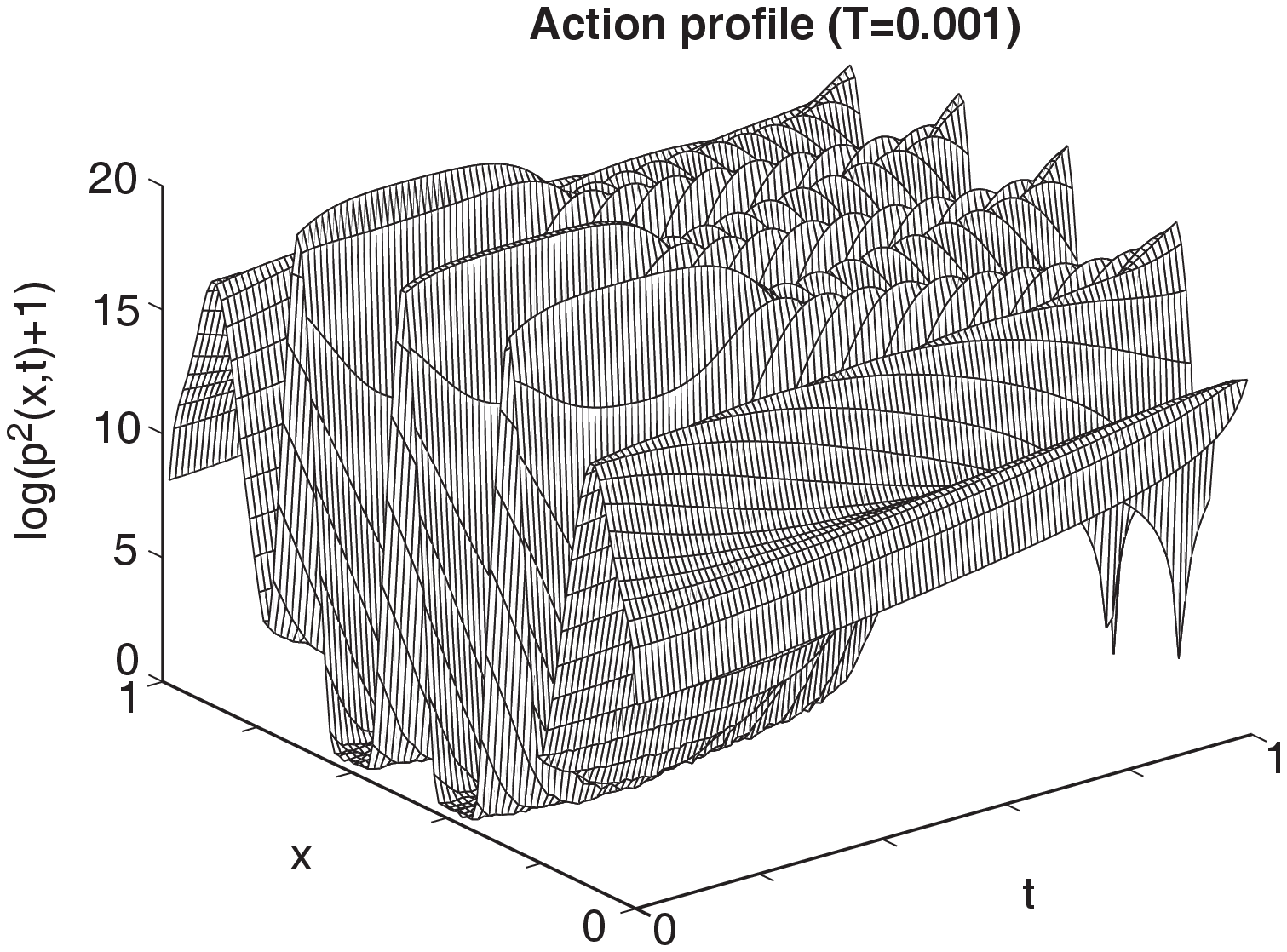}
\caption{We plot the squared noise field $p(x,t)^2$ or action
density as a function of $x$ and $t$ in the case $T=0.001$
(arbitrary units).} \label{fig14}
\end{figure}

In Fig.~\ref{fig10} we show the switching scenario for the
transition times $T=0.001$ for the height, slope and noise. In
Fig.~\ref{fig10} we show the snapshots along the pathway at times
$t= 0.0, 0.00025, 0.00075, 0.001$. In Fig.~\ref{fig14} we depict
the squared noise field or space-time action density. In the short
time regime it is more advantageous to nucleate multiple domain
wall pairs in the slope field, corresponding to multiple steps of
facets in the height field.
\subsection{Switching action}
\begin{figure}
\includegraphics[width=1.0\hsize]{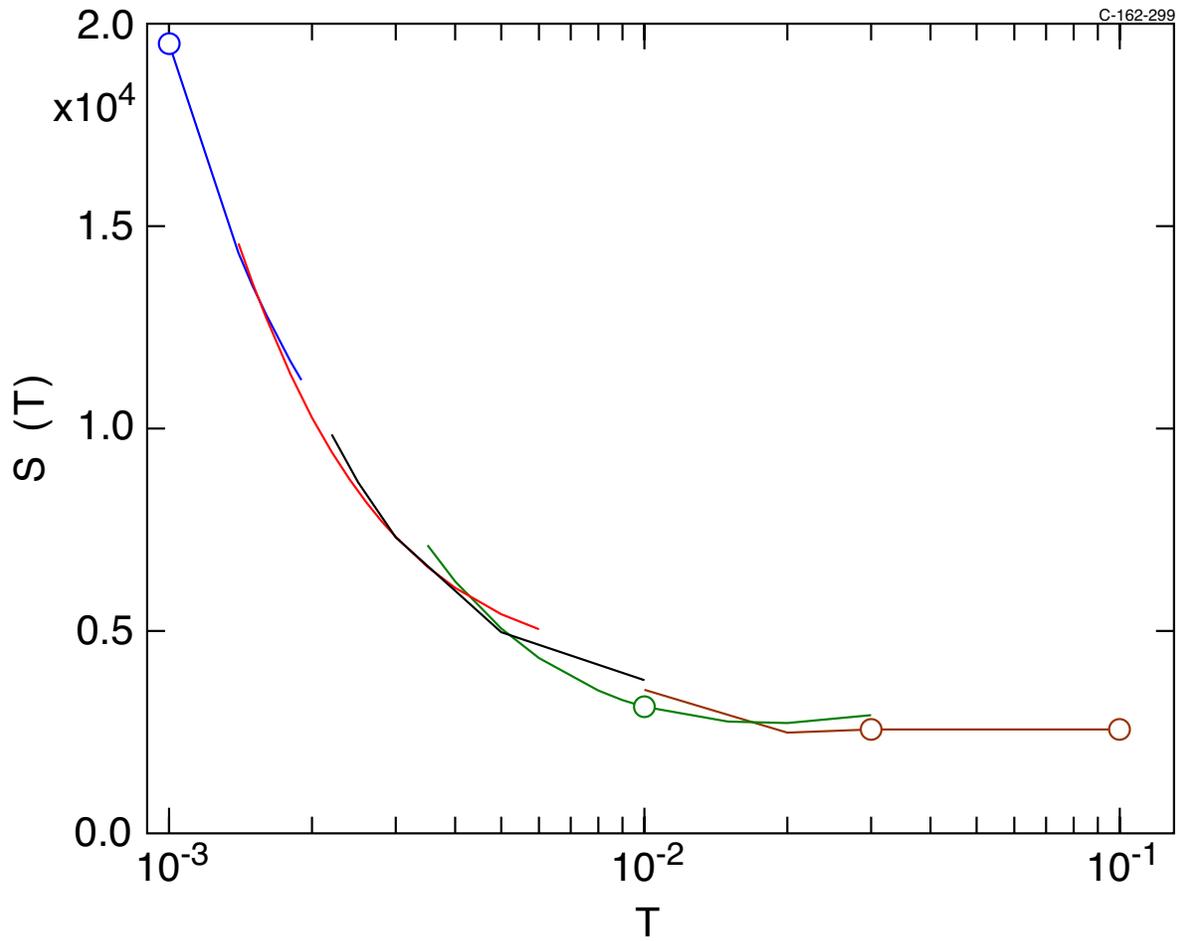}
\caption{We depict the action $S(T)$ as a function of the
transition time $T$ for five transition scenarios. The circles
correspond to the transition pathways for $T=$ 0.01, 0.03, 0.01,
0.001; the remaining pathways involve one nucleation at the center
and one nucleation from the boundary. The plot shows that more
domain wall pairs, yielding a lower action, are nucleated at
shorter transition times.}\label{fig15}
\end{figure}

In Fig.~\ref{fig15} we depict the action $S(T)$ as a function of
the transition time $T$ for five transition scenarios. The circles
correspond to the transition pathways we discussed earlier and
shown in Fig. \ref{fig7}-\ref{fig10} for $T=$ 0.01, 0.03, 0.01,
0.001; the remaining pathways (not shown) involve one nucleation
at the center and one nucleation from the boundary. The plot
clearly indicates that more domain wall pairs, yielding a lower
action,  are nucleated at shorter transition times.

\begin{figure}
\includegraphics[width=1.0\hsize]{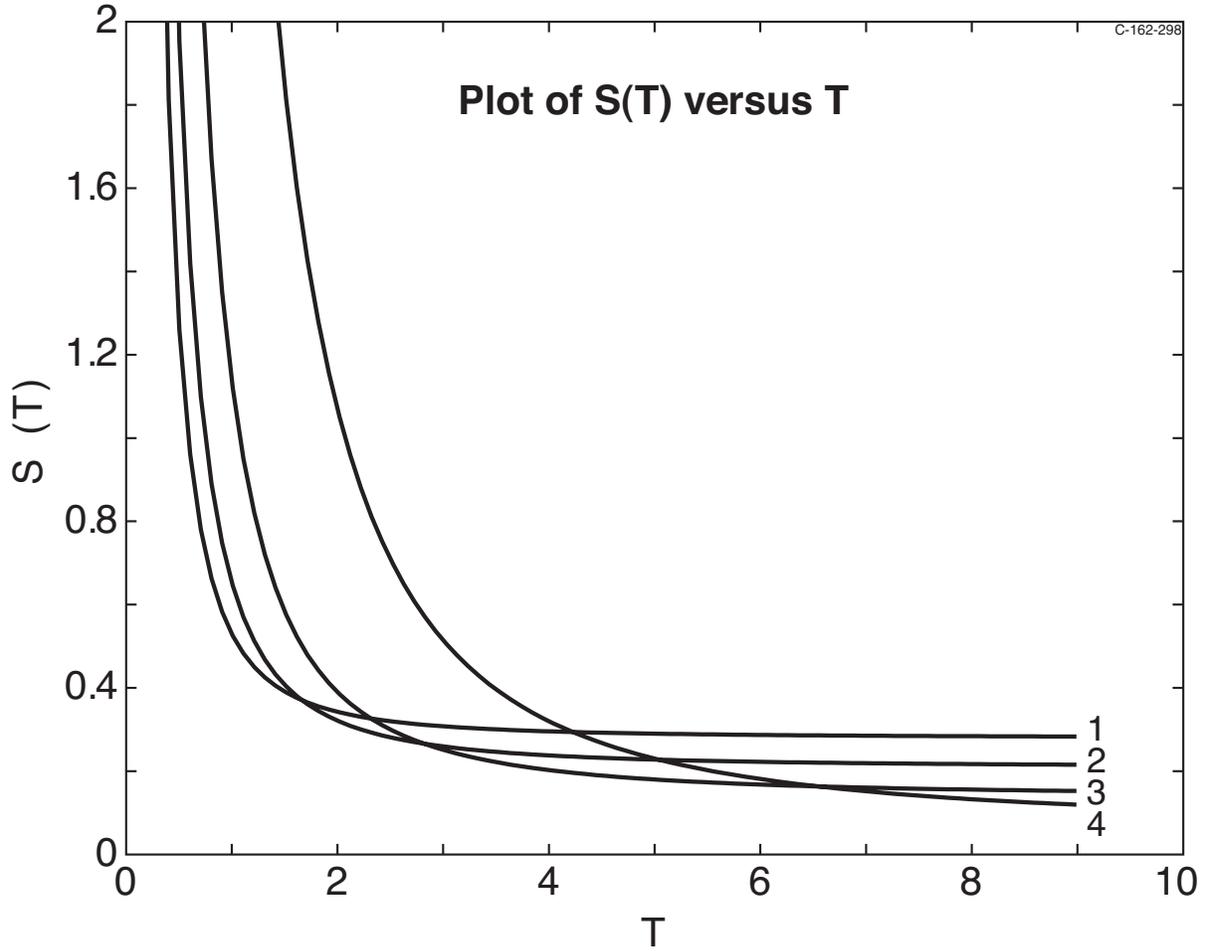}
\caption{ The action $S(T)$ given  by Eq. is plotted as a function
of $T$ for transition pathways involving up to 4 domain walls
pairs in the slope field. The labeling indicates the number of
domain wall pairs. The lowest action and thus the most probable
transition is associated with an increasing number of domain wall
pairs at shorter transition times (arbitrary units).}\label{fig16}
\end{figure}

This relationship can be accounted for by the following
considerations. For a single domain wall pair propagating across
the system the associated action is given by
$S_1=S_{\text{nucl}}+A(L/T)^3T$. Here $S_{\text{nucl}}$ is the
nucleation action associated with the left handed domain wall. The
second term follows from Eq. (\ref{actdw}), where we note that the
velocity $v=L/T$ scales with the amplitude $k$; $A$ is a constant
which we do not have to specify further. In the case of a
transition effectuated by the nucleation and transition of two
domain wall pairs we have correspondingly for the action, $S_2 =
2S_{\text{nucl}}+2A(L/2T)^3T$, where we note that the domain wall
pair only propagate half the distance. In the general case of $n$
domain wall pairs we obtain the expression
\begin{eqnarray}
S_n=nS_{\text{nucl}}+A\frac{L^3}{n^2T^2}. \label{svst}
\end{eqnarray}
In Fig.~\ref{fig16} we have depicted $S(T)$ versus $T$ for
different values of $n$, which shows that the multi domain wall
transitions have lower action at shorter time. This result follows
from the competition between the nucleation action and the action
associated with the propagation and is in qualitative agreement
with the numerical results shown in Fig.~\ref{fig15}.
\section{\label{2d}Transition pathways in 2D}
In 2D the weak noise approach yields elementary spherically
symmetric growth modes. In terms of the diffusive field $w$ the
diffusion equation (\ref{diff}), $\nabla^2w=k^2w$, has the
asymptotic growing solution $w_+\propto \exp(kr)$ for $r\gg 1/k$
giving rise to the height field $h_+=(k/k_0)r$ and the slope field
${\bf u}_+ =(k/k_0){\bf r}/r$. Likewise, the non-linear
Schr\"odinger equation (\ref{nls}), $\nabla^2w=k^2w-k_0^2w^3$,
yields the decaying solution $w_-\propto \exp(-kr)$ and,
correspondingly, $h_- = -(k/k_0)r$ and ${\bf u}_-= -(k/k_0){\bf
r}/r$. The height modes correspond to a tip (upward cone) and a dip
(downward cone) in the interface profile, whereas the slope modes
are outward and inward pointing vector fields of constant magnitude
$k/k_0$, i.e., monopole fields. Like in the 1D case the static
growth modes can be boosted to a finite propagation velocity and one
can construct a dynamic growth morphology in terms of a dilute gas
of monopoles in the slope field with superimposed diffusive modes.
In a charge language the positive monopoles are solutions of the
noiseless Burgers equation and carry no action, whereas the negative
monopoles carry an action $S\propto (\nu^2T/k_0^2)k^2$. In order to
model a pathway from $h_{\text{init}}$ at $t=0$ to $h_{\text{fin}}$
at $t=T$ one assigns a gas of monopoles representing
$h_{\text{init}}$. With the appropriate assignment of the noise
field corresponding to nucleation events this configuration will
evolve in time to $h_{\text{fin}}$. The total action associated with
negative growth modes, using the WKB ansatz
$P\propto\exp(-S/\Delta)$, then yields the transition probability
for the kinetic pathway. Details of this procedure has been
discussed at length in Ref. \cite{Fogedby06a} and will not be
reproduced here.

\begin{figure}
\includegraphics[width=0.6\hsize]{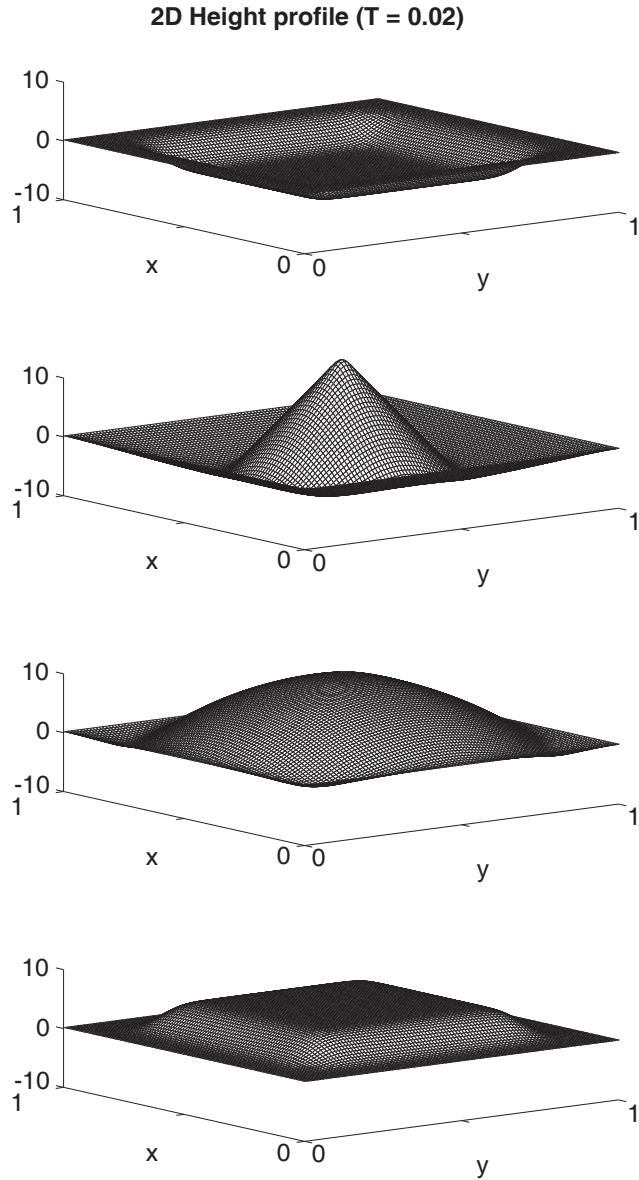}
\caption{We depict a 2D long time transition scenario for the
height profile from an initial plateau at $h=-5$ to a final
plateau at $h=+5$. The transition time is $T=0.02$. The transition
takes place subject to the nucleation of a single peak in $h$ at
the center. The peak eventually broadens as we approach the final
configuration (arbitrary units).}\label{fig17}
\end{figure}

\begin{figure}
\includegraphics[width=0.6\hsize]{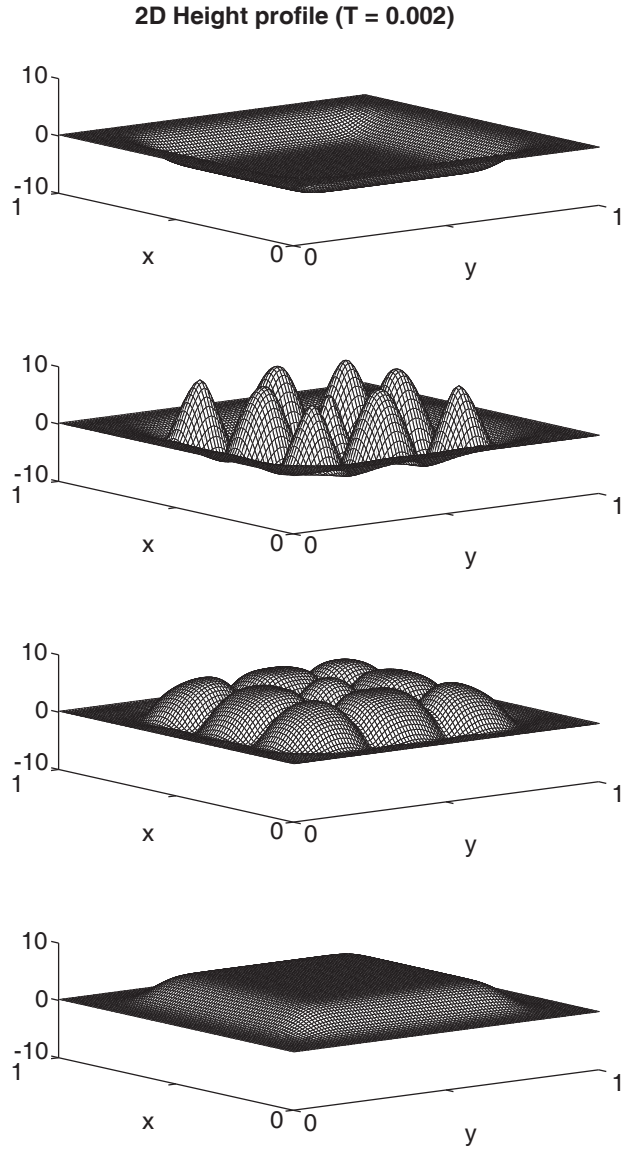}
\caption{We depict a 2D short time transition scenario for the
height profile from an initial plateau at $h=-5$ to a final
plateau at $h=+5$. The transition time is $T=0.002$. The
transition takes place subject to a regular pattern of 9
nucleation zones. The peaks eventually broaden as we approach the
final configuration (arbitrary units).}\label{fig18}
\end{figure}

The minimum action method is easily extended to higher
dimension generalizing the procedure in Sec. \ref{num}. Choosing
the parameters $\nu=1$, $\lambda=2$, and a $100\times 100\times
100$ set of $xyt$ grid points and matching the height profile to
the boundary values $h({\bf r})=0$ by a 2D generalization of Eqs.
(\ref{leftinit}-\ref{rightinit}), we have in Figs~\ref{fig17} and
\ref{fig18} depicted the 2D switching scenarios for the height
field  at transition times $T=0.02$ and $T=0.002$ from an initial
plateau at $h=-5$ to a final plateau at $h=5$. In the case
$T=0.02$ a single peak in $h$ is nucleated at the center of the
plateau $h=-5$. The peak amplitude evolves in time and eventually
flattens to the plateau at $h=5$. In the case $T=0.002$ the
transition takes place subject to the nucleation of a regular
pattern of growing cones in $h$ which eventually broadens and
merge together. Like in the 1D case we note again that more peaks
are nucleated at shorter transition times.
\section{\label{sum}Summary and conclusion}
In the present paper we have applied the minimum action method
based on the Freidlin-Wentzel scheme for rare events driven by weak noise
to the KPZ equation for a growing interface. The KPZ equation is a
non-gradient system and a characterization of kinetic pathways in
a free energy landscape is not available, on the other hand, the
pathways can be characterized as taking place in an action
landscape. Correspondingly, the transition probabilities are
characterized by the associated action of a specific pathway,
unlike the free energy case for gradient systems where free
energy considerations apply in the evaluation of the Arrhenius
factor for the transition.

The minimum action method basically identifies the
kinetic pathway in the action landscape by seeking a minimum of
the action using an optimization technique. Once the minimum has
been reached the method provides the kinetic pathway subject to
given initial and final configurations combined with appropriate
boundary conditions. We have conducted a detailed analysis of the
1D case and find that the pathways can be characterized by the
nucleation and subsequent ballistic propagation of growth modes.
These growth modes correspond to moving facets or steps in the KPZ
height field and to moving domain walls in the slope field. We
also find that the numerical results are in good qualitative
agreement with the canonical phase analysis previously developed
for the KPZ equation. We have, moreover, applied the minimum
action method to the 2D case.

In conclusion, we believe that the minimum action method
provides a new tool in analyzing the kinetics of spatially
extended or field theoretical non-gradient systems like the KPZ
equation studied here. The method supplements previous scaling
analysis of the KPZ equation in focussing on the pattern formation
or many body aspects of kinetic transitions in the weak noise
limit.

\acknowledgments

The work of H. Fogedby has been supported by the Danish Natural
Science Research Council under grant no. 95093801. The work of W.
Ren is supported by NSF grant DMS-0806401 and the Sloan
fellowship. Discussions with A. Svane are gratefully acknowledged.

\newpage
\end{document}